\documentclass[12pt,a4paper]{report}

\usepackage{amsmath}
\usepackage[dvips]{graphics}
\usepackage[dvips]{graphicx}
\usepackage{lscape,longtable}

\newcommand{\sz}{\scriptsize}
\newcommand{\gap}{\vspace{0.6cm}}

\title{\bf{Morphological Classification of Galaxies Using Artificial Neural Networks}}

\author{Nicholas M. Ball \\ \\ Supervisor: Dr Jon Loveday \\ Submitted for the degree of M.Sc. in Astronomy \\ University of Sussex}

\begin{document}

\maketitle

I have read and understood the definition of ``plagiarism'' as set out
in the \textbf{General Assessment Handbook for Masters and Postgraduate
Diploma Candidates} under section 11.

This dissertation is my own work, except where explicitly stated.

\vspace{3cm}

Signed..............................\hspace{3cm}  Date...............

\vspace{1.6cm}

Nick Ball

\tableofcontents
\listoffigures
\listoftables



\chapter{Introduction}

The purpose of this project is to compare the accuracy and consistency of human eyeball-based methods of morphologically classifying galaxies with that of automatic methods. In particular, galaxy types assigned by humans using the well known Hubble 'Tuning Fork' system are compared with corresponding types output by artificial neural network (ANN) programs.

What is the point of classifying galaxies? Of course, galaxy classification is not an end in itself, but a first step towards a greater understanding of the physics of galaxies. Many examples exist in science where a good classification system has led to a much improved understanding. A classic example is the periodic table of the elements: the ordering of elements by their observable properties showed patterns which led to the models of the structure of the atom, and to predictions of new elements which were subsequently discovered. Another example is the classification of lifeforms into species. If a classification system for an observed phenomenon is based on continuously varying parameters and those parameters are in turn shown to be important to the theories which attempt to explain the phenomenon then the classification system helps quantify the connection between the theory and the observations.

The study of galaxies, their formation and evolution is still a science which is in its infancy, which is why people are still trying to perform basic classification. The youth of the science is illustrated by the fact that before the mid 1920s it was not even known for sure whether galaxies are separate systems outside our own Milky Way, or whether they were simply another type of nebula in our own galaxy. Although the idea had been suggested by Immanuel Kant in 1755, and the spiral nebulae clearly observed by Lord Rosse in the mid nineteenth century, it wasn't until such events as the 'Great Debate' in 1920, the discovery by Edwin Hubble of Cepheid variable stars in the Andromeda galaxy in 1923, giving a measure of its distance which vastly exceeded previous estimates, and Hubble's discovery of the expansion of the universe in 1929 that it was accepted that galaxies are indeed separate 'island universes', comparable in size to our galaxy.

Once it was realised that galaxies are indeed separate systems, a logical step was to try to find patterns in their properties. Of course, the 'nebulae' had already been studied in detail since Lord Rosse's time, but nebulae in our galaxy were also included in the same studies. In the 1920s astronomy was confined to the optical waveband (radio astronomy did not begin until the 1930s and other wavebands were not exploited until even later), so the properties of galaxies that were studied were their visual appearance and their spectra. It was clear that galaxies came in different types, i.e. elliptical, spiral and irregular, and much further detail within their structure was visible in nearby examples, so it was inevitable that some sort of classification system would be set up. What is remarkable is that the system that was set up by Hubble in 1926, though only based on the appearances of these bright nearby galaxies, has proven to be so useful. No clearly superior system to describe observed galaxies has yet been developed.

For the classification to be meaningful, the parameters must be physically motivated, or be shown to be correlated in a way which might be explainable by a physical model. A galaxy's appearance, spectrum etc. could simply be described, and patterns could be found, but the idea is to find a physical model which \emph{explains} the patterns. Ultimately a complete theory should mean that any classification sysem is entirely objective.

A good analogy for the ideal galaxy system is the Hertzsprung-Russell diagram used in stellar astronomy. This plots stellar absolute magnitude against colour (or equivalently luminosity against temperature) and shows well-known correlations such as the main sequence and the red giant branches. These have since been explained by detailed models of stellar evolution. The hope is that an analogous discovery process can occur for galaxies, to explain their formation and evolution into the forms seen today. Another motivation for classifying galaxies is to provide catalogues of objects for further study, and, at low redshift, a comprehensive base from which comparisons can be made with objects as higher redshift, where properties such as detailed morphology cannot be observed.

The structure of this thesis is as follows: in Chapter 2 important classification systems are reviewed. They are important because several independent human experts are trained in and can therefore use these systems to classify galaxies. They provide the basis for comparisons between humans and automated systems, the aim of the project. In particular the de Vaucouleurs 'T' system, based on the Hubble system, has been found to be useful. Chapter 3 explains the increasing amount of data available to astronomers, the consequent need for automated classifications and why ANNs are the best solution. ANNs are then described, concentrating on the concepts which are important in their application to this project. This is followed by the results of the comparisons in Chapter 4. No human classifications are undertaken in the project, but instead a number of neural networks are constructed and their outputs compared to various sets of eyeball classifications and human-ANN comparisons published in the literature. In particular, the eyeball classifications of Shimasaku et al. (2001), who classify 456 galaxies from the Sloan Digital Sky Survey (SDSS) commissioning data, are used. The networks are also applied to the SDSS Early Data Release. Possible extensions to the work, including the important extension to classifying galaxies spectrally, are described in Chapter 5, followed by conclusions. A general listing for the network programs is included as Appendix B.

\chapter{Classification Systems}

\section{The Hubble System}

By far the most well-known classification system, the Hubble (or Mount Wilson) system was first devised by Edwin Hubble (Hubble 1926), and has remained essentially the same to the present. Some subsequent modifications were made by him (Hubble 1936) and the system is given its definitive description by Sandage in the Hubble Atlas of Galaxies (Sandage 1961). It is most recently fully described in The Carnegie Atlas of Galaxies (Sandage \& Bedke 1994). The system divides nearby bright galaxies in the visual waveband into ellipticals, lenticulars, spirals, barred spirals and irregulars to give the famous Tuning Fork diagram (figure 2.1).

\begin{figure}[!htbp]
\begin{center}
\includegraphics[angle=270, width=12cm]{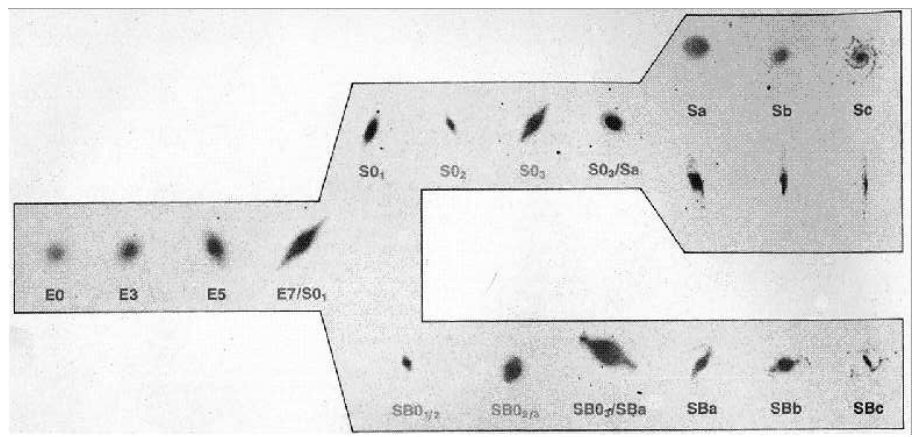}
\caption[The Hubble Tuning Fork diagram]{The Hubble Tuning Fork Diagram, including examples of each main galaxy type (from Gene Smith's Astronomy Tutorial, {\tt{http://casswww.ucsd.edu/public/tutorial/Galaxies.html}})}
\end{center}
\end{figure}

In the original 1926 system the parameters are, for ellipticals, the ellipticity $\epsilon$, given by $\epsilon=10(a-b)/a$ where $0\le\epsilon\le7$ and $a$ and $b$ are the projected major and minor axes on the sky; for spirals both the central concentration of light and the tightness of the spiral arms define the sequence Sa--Sb--Sc. Type Sa galaxies have large nuclei and strongly wound smooth spiral arms. Type Sc have small nuclei and loosely wound arms which are generally more patchy, i.e. resolved into stars and HII (ionised hydrogen) regions. Sb's are intermediate. The presence or absence of a bar gives the other prong of the tuning fork, and the barred galaxies are divided in a similar way into Sba--Sbc. Irregular galaxies, i.e. those with no obvious structure, are given as a separate class. In 1936, S0 and SB0 (lenticular and barred lenticular) galaxies were added (Hubble 1936) and suggested as approximately a transition type between E7 and Sa/SBa. The nature of this transition is still controversial. The low resolution of Sc--Irr was given finer division by later workers and in 1959 became the main axis of the de Vaucouleurs three dimensional system (\S 2.3), consisting of E0--S0--Sa--Sb--Sc--Sd--Sm--Im--Irr, with finer subdivisions ${\textnormal{S0}}^{-,0,+}$, Sab, etc. (de Vaucouleurs 1959). This system is also known as the Revised Hubble System. Type Sd is what would previously have been designated late Sc (i.e. nearer to Irr than to Sb), and Sm and Im are further subdivisions reflecting the division of irregular galaxies into barred and unbarred forms (e.g. the Large Magellanic Cloud has a weak bar structure and is thus of type Im; Im stands for Irregular Magellanic).

Although only based on the optical appearance of bright nearby galaxies, the Hubble Sequence has proven to correlate very well with many more general physical parameters of these galaxies. Combined with the fact that no clearly superior classification system has been found, this is why the Hubble system is so well-known and extensively used today. Examples of correlations include (from Lahav et al. 1995), integrated colour, dynamical properties such as stellar velocity dispersions and rotation curves, mass in free neutral hydrogen (HI) and, more broadly, galaxy overall mass and luminosity (see table 2.1). The mean colour is particularly significant because the integral of this for a galaxy reflects the mean spectral type of its stellar population. This connects with theories of galaxy formation and evolution, which predict galaxy properties such as their dynamics and star formation history. Further important parameters are the concentration parameter and the bulge to disk ratio, which measure the relative concentration of light towards the centre of the galaxy. Elliptical galaxies are diskless and the ratio changes through to Sc galaxies and beyond which appear bulgeless. The bulges are dominated by old red stars and the disks by young blue stars. Thus the bulge to disk ratio and concentration parameter correlate with Hubble type and overall galaxy colour. The concentration parameter is described in Chapter 4.

\begin{landscape}
\begin{table}[p]
\begin{center}
\caption[Galaxy characteristics correlate with classification]{Galaxy characteristics correlate with classification (from Gene Smith's Astronomy Tutorial, {\tt{http://casswww.ucsd.edu/public/tutorial/Galaxies.html}}).}
\gap
\begin{tabular}{|c|c|c|c|c|c|c|}
\hline
                             &\sz{E0-E7}                         &\sz{S0}            &\sz{Sa}           &\sz{Sb}     &\sz{Sc}          &\sz{Irr}                        \\
\hline \hline
\sz{Nuclear Bulge}           &\sz{"All bulge", no disk}          &\sz{Bulge \& disk} &\sz{Large}        &            &\sz{Small}       &\sz{None}                       \\
\sz{Spiral Arms}             &\sz{None}                          &\sz{None}          &\sz{Tight/smooth} &            &\sz{Open/clumpy} &\sz{Occasional traces}          \\ 
\sz{Gas (mass)}              &\sz{Almost none}                   &\sz{Almost none}   &\sz{\verb+~+1\%}  &\sz{2--5\%} &\sz{5--10\%}     &\sz{10--50\%}                   \\
\sz{Young Stars HII Regions} &\sz{None}                          &\sz{None}          &\sz{Traces}       &            &\sz{Lots}        &\sz{Dominates appearance}       \\
\sz{Stars}                   &\sz{All Old(\verb+~+$10^{10}$ yr)} &\sz{Old}           &\sz{Some young}   &            &                 &\sz{Mostly young, but some old} \\ 
\sz{Spectral Type}           &\sz{G-K}                           &\sz{G-K}           &\sz{G-K}          &\sz{F-K}    &\sz{A-F}         &\sz{A-F}                        \\
\sz{Color}                   &\sz{Red}                           &\sz{Red}           &                  &            &                 &\sz{Blue}                       \\
\hline
\sz{Mass (M)}                &\sz{$10^{8}$--$10^{13}$}           &\multicolumn{4}{c|}{\sz{(More)$10^{12}$--$10^{9}$(Less)}}            &\sz{$10^{8}$--$10^{11}$}        \\ 
\sz{Luminosity (L)}          &\sz{$10^{6}$--$10^{11}$}           &\multicolumn{4}{c|}{\sz{(More)$10^{11}$--$10^{8}$(Less)}}            &\sz{$10^{8}$--$10^{11}$}        \\
\hline
\end{tabular}
\end{center}
\end{table}
\end{landscape}

However, the Hubble system is by no means perfect. It does not form a complete system for describing all galaxies, or indeed most in numerical terms, since most galaxies are small and faint, like the majority of stars. Some parameters are mixed up along the length of the sequence, for example bulge size, spiral arm smoothness and tightness all change from Sa to Sc as described above. The ellipticity $\epsilon$ is a function of projection and true galaxy triaxiality. The system shows no correlation with disk galaxy morphological properties in the near-infrared (Block et al. 1999), where dust obscuration is around 10\% of that in the optical. Block et al. describe the dust in the visual as a 'mask' of insignificant dynamical mass which obscures galaxy disks. The Hubble type therefore does not correlate with the dynamical mass distribution in spiral galaxy disks in the infrared. The system does not explain the forms of galaxies at moderate and high redshift, i.e. $z\ge 0.1$, where galaxy evolution becomes significant. Further problems include the fact that galaxies in rich clusters are not well described, being mostly S0 or elliptical and thus the types are poorly resolved. All types of irregular galaxies are lumped together. Low surface brightness galaxies, dwarf spheroidals and spirals, and other unusual types such as amorphous galaxies (e.g. M82) or cD galaxies (see \S 2.2), are not described by the system (van den Bergh 1998). Active galaxies (AGN, quasars) are also not described, although these only form a small percentage of galaxies in the local universe. Those undergoing mergers are often recognisable as Hubble types, but are hard to quantify as such. Many of these problems are because the system was not designed to address them, but this does not alter its incompleteness.

The Hubble system is important in this project because it forms the main axis of the de Vaucouleurs 3D system which in turn maps directly onto his numerical one-parameter 'T' system, described in \S 2.3.1. This is the system used for the comparisons between humans and artificial neural networks here and in the literature. Indeed, it is the only system which \emph{could} practically be used for such a comparison, since it is the only one many independent experts are familiar with, and there is no obviously better system to learn and use. It is therefore worth making the comparisons in spite of the fact that it will not be the sought after final objective method of classifying galaxies.

\section{The Yerkes System}

This system, devised by W.W. Morgan (1958) is important because it uses the central concentration of light in a galaxy as its fundamental parameter. This has been shown to correlate with the mean stellar spectral type in galaxies, and in spiral galaxies reflects the ratio of the bulge (old red stars) to the disk (light dominated by young blue stars). The radii at which certain percentages of the total flux are received from the galaxy are found, and their ratios taken (exact definitions of the concentration index can vary, and the ones used in this project are given in Chapter 4). Like the T parameter, the index is one dimensional, and is thus useful in the automated classification systems used today. For example, the ratio $r_{50}/r_{90}$ (50\%/90\% of flux) is commonly used (again see Chapter 4), and is employed in this project and in the literature. The parameter is also useful because it can be measured for galaxies to a much greater distance (smaller image) than can any detailed morphology. 

In the Yerkes system, the galaxy types are designated a--af--f--fg--gk--k in order of increasing concentration (a=spiral type Sc or irregular, k=elliptical). Secondary 'form families' are added (ESBILND, where E=elliptical, S=spiral, B=barred, I=irregular, L=low surface brightness, N=bright or active nuclei, D=rotationally symmetric), and a number is added to give the flattening of the galaxy. A later form addition is a subclass of particularly luminous D galaxies called cD, giving the well known cD type of huge spherical galaxies found in the centres of rich galaxy clusters. cD's do not fit in the Hubble system.

\section{de Vaucouleurs' Extension to the Hubble System}

Described in de Vaucouleurs (1959), and illustrated in figure 2.2 this system is also known as the Revised Hubble system. It extends the tuning fork to three dimensions, and has orthogonal axes of the Hubble stage (E--S0-- ... --Sd--Im), the 'family' SA--SB (ordinary-barred), and 'variety' S(s)--S(r) (s-shaped to ringed). The Hubble stage axis maps onto the T system (de Vaucouleurs et al. 1991). This is reasonable because is has been shown that the family and variety axes have a much less significant correlation with physical properties than the stage axis (van den Bergh 1998). The stage axis is not perfect, because at the spiral-irregular end luminosity and colour effects are mixed up and mapped onto the single parameter (van den Bergh 1998). \\

\begin{figure}[!htbp]
\begin{center}
\includegraphics[angle=270, width=12cm]{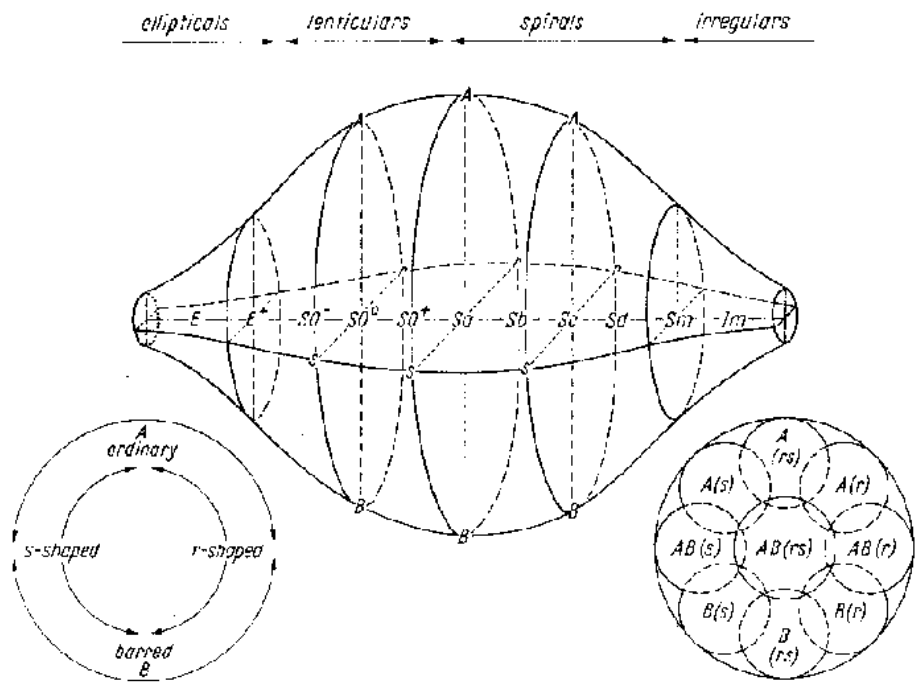}
\caption[The de Vaucouleurs 3D system]{The de Vaucouleurs 3D system (de Vaucouleurs 1959). The size of the family and variety axes reflect the extent to which the features are present at that Hubble stage. The family and variety axes are not thought to be significant in terms of the physics of galaxy formation and evolution.}
\end{center}
\end{figure}

\subsection{The de Vaucouleurs T System}

This is the system used in the comparisons in this project. It has turned out to be a useful system because it is a one-to-one mapping between the letter-based Hubble stages, which are the one system that many different experts have experience with, and the integer T types, which can be output by a neural network as a classification (either as integers or a continuous sequence). The system was used in the only extensive study carried out for comparison between humans and between humans and artificial neural networks (Naim et al. 1995a,b, Lahav et al. 1995, 1996). As mentioned above, the parameter is not perfect, but for comparison with classifications by several humans it is the only practical solution. The Hubble and T systems are shown in table 2.2.

\begin{table}[!htbp]
\begin{center}
\caption[Hubble type versus T type in the de Vaucouleurs 3D system]{Hubble type versus T type in de Vaucouleurs' extension to the Hubble system. N.B.: cE = compact elliptical, cI = compact irregular, I0 = irregular: non-Magellanic/amorphous, Pec = peculiar/unidentified}
\gap
\begin{tabular}{|c|cccccccccc|}
\hline
T Type      &-6  &-5 &-4                 &-3                  &-2                  &-1                  &0    &1  &2   &3   \\
\hline
Hubble Type &cE  &E0 &$\textnormal{E}^+$ &$\textnormal{S0}^-$ &$\textnormal{S0}^0$ &$\textnormal{S0}^+$ &S0/a &Sa &Sab &Sb  \\
\hline \hline
T Type      &4   &5  &6                  &7                   &8                   &9                   &10   &11 &90  &99  \\
\hline
Hubble Type &Sbc &Sc &Scd                &Sd                  &Sdm                 &Sm                  &Im   &cI &I0  &Pec \\
\hline
\end{tabular}
\end{center}
\end{table}

\section{Other Systems}

Many other systems have been devised. Some are purely descriptive, and therefore not claimed to be actual \emph{classification} systems. They can be very detailed (examples are Wolf (1908), which includes nebulae and the Vorontsov-Velyamov et al. (1962) Morphological Catalogue of Galaxies). Others only cover certain types of galaxies, or simply extend the Hubble sequence, for example Van den Bergh (1960) on spiral arm luminosity classes, although this covers such aspects as the correlation between arm type ('grand design' or flocculent) and luminosity, which the Hubble system does not correlate with. Still other systems have been superceded. Parameters such as the central concentration of light and, perhaps less usefully in hindsight, many very detailed aspects of optical spiral arm morphology, turn up in many of these systems. The spiral arm morphology details will probably be more useful when detailed galaxy formation and dynamics are better understood.

Another important set of systems are those using purely spectral parameters. As mentioned, the integrated spectrum of a galaxy reflects its stellar population and galaxies of similar morphological type tend to have similar stellar populations. Hence spectral and morphological types correlate quite well. As with concentration index, spectral types can be measured to great distances. The first examples are Humason (1936) and Morgan \& Mayall (1957), with many more since (e.g. Folkes et al. 1996, which also uses artificial neural networks). \\

More details of all the systems described in this chapter, and of further systems, are given in van den Bergh (1998) and the history by Sandage (1975).

However, no matter how good the system, it is still of very limited use if it requires a human to manually classify each individual galaxy, because modern sky surveys produce data on orders of magnitude too many galaxies. Automatic systems are needed and this is described in the next chapter.


\chapter{Artificial Neural Networks (ANNs)}

\section{The Need For Automation}

\noindent There are two reasons why automatic classification of galaxies is desirable:

\begin{itemize}
\item The amount of data produced by modern sky surveys is simply too large for humans to classify manually
\item Humans are not objective and consistent classifiers
\end{itemize}

\subsection{The Increasing Amount of Data}

Increasing data, both galaxy images and masses of accompanying parameters, is becoming available in quantities vastly too large to classify manually. The increase is essentially technology driven, for example CCDs have the advantage of linear response to light intensity, and the data is directly stored digitally. They are also up to 90\% efficient at collecting photons of light, as opposed to the few percent efficiency of photographic plates. These attributes, plus the exponentially increasing computer processing power available, allow the larger datasets to be of more uniform properties and of higher quality than potentially variable photographic plates. 

Table 3.1 illustrates the increasing amount of data which has become available, with some landmark surveys. The number of galaxies surveyed and the number of redshifts available (which give the distance to a galaxy) are given.

\begin{landscape}
\begin{table}[p]
\begin{center}
\caption[Some important optical surveys to illustrate the increasing amount of data available]{Some important optical surveys to illustrate the increasing amount of data available. Note: [1] Many of the nebulae are galaxies, others are e.g. star clusters, planetaries; the catalogue was extended from 103 to 110 later. [2] Again many of which are galaxies - note that the external nature of galaxies was not agreed upon until the late 1920s. References: 1771--1967 - Seitter (1987), Fairall (1997); RC1 - de Vaucouleurs et al. 1964; Cfa1 - Davis et al. 1982; APM - Maddox et al. 1990; LCRS - Shectman et al. 1996; 2dF - Colless et al. 2001 (astro-ph 0106498); SDSS - York et al. 2000; VISTA - http://www.vista.ac.uk}
\gap
{\footnotesize
\begin{tabular}{|l|l|l|l|l|}
\hline
    Date(s)      &    Survey                                                   &    A.k.a.   &    No. galaxy images  &    No. redshifts \\
\hline \hline
    1771         &    Messier - Messier Catalogue                              &    M1-M103  &    103 nebulae [1]    &                  \\
    1888         &    Dreyer - New General Catalogue                           &    NGC      &    4630+ objects [2]  &                  \\
    1914         &    Slipher - taking galaxy spectra                          &             &                       &    13            \\
    1934         &    Hubble (in the Realm of the Nebulae)                     &             &    c. 44,000          &    100+          \\
    1956         &    Humason, Mayall \& Sandage                               &             &                       &    800+          \\
    1964         &    Reference Catalogue of Bright Galaxies                   &    RC1      &                       &    $<$ 1,500     \\
    1967         &    Lick Observatory survey on the distribution of galaxies  &             &    1,000,000+         &                  \\
    1982         &    Centre for Astrophysics Redshift Survey                  &    Cfa1     &                       &    2,437         \\
    1985--1990   &    Automatic Plate Measuring machine Galaxy Survey          &    APM      &    c. 2,000,000       &                  \\
    1988--1994   &    Las Campanas Redshift Survey                             &    LCRS     &                       &    26,418        \\
    1996--2001   &    Anglo Australian Observatory 2 degree field              &    2dfGRS   &    Based on APM       &    c. 250,000    \\
                 &    Galaxy Redshift Survey                                   &             &    (467,214)          &                  \\
    2000--2004   &    Sloan Digital Sky Survey                                 &    SDSS     &    c. 50,000,000      &    c. 1,000,000  \\
    2004--2016+  &    Visible and Infrared Survey Telescope for Astronomy      &    VISTA    &    50Tb of images     &                  \\
\hline
\end{tabular}
}
\end{center}
\end{table}
\end{landscape}

Classifying galaxies manually takes time, because a human expert must look at each individual image and decide which class the galaxy belongs to, or if it is uncertain. Examples of the largest manually classified catalogues are the Third Reference Catalogue of bright Galaxies (RC3, de Vaucouleurs et al. 1991), with nearly 18,000 galaxies, and the European Southern Observatory photometric catalogue ESO-LV (Lauberts \& Valentijn 1989) with more than 15,000 (numbers from Naim et al. 1995a). These took several years to compile. The new surveys (see table) will have images of up to 50 million galaxies, therefore an automatic classification system is vital. The system must be reliable, i.e. classify with known, quantified and acceptably small errors, and fast, so that the millions of galaxies can be processed on a reasonable timescale. Once the image parameters have been determined (also by automatic programs), automated algorithms can classify hundreds of galaxies per second, so millions of galaxies would take a few hours. The table also shows that even data from the 1930s has not been fully classified and parameterised as it could be, so much of the datasets already available remain largely unexplored.

A significant recent dataset is the Sloan Digital Sky Survey Early Data Release (SDSS EDR, Stoughton et al. 2001). This includes over 120 parameters for each of 13,804,448 objects, around half of which are galaxies. In this project, simple networks were applied to a subset of this data (see Chapter 4).

\subsection{Human Objectivity}

Automatic systems are also desired because they are more objective, i.e. not subject to the conscious and unconscious prejudices which affect humans in looking at galaxy images, no matter how well intentioned. For example someone who has always used the same system is biased towards looking for parameters which define the system, even though these parameters may be essentially arbitrary or totally empirical. Another example is the fact that the human brain has evolved to see patterns very easily, and sometimes see patterns which are not really there (the man who saw the tiger that wasn't there survived while the one who didn't see the one that \emph{was} there did not). So spurious patterns in images could easily be picked up upon. Automatic systems, while by no means perfect, are at least more quantifiable, and although extraction of features from images is by no means a simple problem, it can be done reliably, and quantitatively.

\subsection{The Solution}

Objective classifiers can be constructed from conventional statistics, such as principal component analysis (PCA), but these are limited to being linear, and thus have to make linear approximations. Artificial neural networks (ANNs) allow the implementation of nonlinear statistics, which better reflect the underlying distribution of galaxy parameters. In combination with the increased consistency and size of the new digital sky surveys becoming available, as described above, much new information can be found. ANNs are now described.

\section{Introducing ANNs}

In this section artificial neural networks (ANNs) are introduced. The networks used for classifying galaxies involve one or more layers of neurons, with the galaxy parameters as the input to the first layer and the classification as the output of the final layer. The best networks from previous studies used multiple layers. How ANNs work is described by beginning with a single neuron, generalising to a layer of neurons and then to many layers. The expressions used retain their form at each stage, but the scalars for a single neuron with a single input become vectors and matrices for the more complex arrangements. Additional terms such as weighting between layers also appear. Many other types of neural network besides the multilayer type exist, but these are not described in detail, because they have not been used in human--ANN comparisons. This does not mean that they could not be used, however. Some examples are given in the following. See also Bishop (1995), or the M{\sc{atlab}} documentation at {\tt{http://www.mathworks.com/access/helpdesk/help/toolbox/nnet/nnet\\.shtml}}.

Artificial Neural Networks (ANNs) are structures which mimick the neural structure of the brain, in the sense that they consist of neurons acting as nodes which are able to carry out processing. The neurons are interlinked by one or more connections to other neurons. The connections can be weighted to alter the output from one neuron before it is input to the next. Whilst the networks may mimic the brain in this basic sense, they in no way approach the same level of complexity. The human brain has of order $10^{11}$ neurons, each with a few to a few thousand connections, but the networks used for classifying galaxies only have up to hundreds of neurons, in a few layers with of the order of ten inputs and outputs, and the necessary number of weights. Thus it is more appropriate to think of them acting as a reasonably complex nonlinear mapping than a 'miniature brain'.

The purpose of a classifier is to map a set of inputs onto the correct output (or outputs), the inputs being the parameters for the object in question and the output the classification. The classifier is trained on examples, or given a criterion for grouping data, and after this should be able to cope with new examples which it has not seen before. In the case of galaxies (and in most cases) the mapping is not simple but is \emph{nonlinear}. ANNs are used because they are able to incorporate this nonlinearity, whereas conventional statistics are not. Linear classifiers do work, to some extent, but neural networks are better (e.g. Naim et al. 1995b). Examples of linear classifiers are principal component analysis for extracting key parameters (see \S 3.7), or the Na\"\i ve Bayes classifier, which uses basic Bayesian probabilities (Bazell \& Aha 2001).

As well as classifying galaxies, ANNs have been found useful for various other applications in astronomy. Examples are star--galaxy separation on images (determining whether an object is a star or a galaxy), stellar spectral classification, adaptive optics, scheduling observation time and time dependent applications such as predicting solar activity (reviews are Miller 1993 and Storrie-Lombardi \& Lahav 1994).

\section{Single Neuron}

\subsection{One Input}

The simplest neuron has one {\bf{input}}, one {\bf{output}}, a {\bf{weight}}, and some kind of {\bf{transfer function}} which operates on the input to give the output. Also present is a {\bf{bias}}, analogous to that in a DC electric circuit. The input is multiplied by the weight, the bias is added, and the transfer function operates on the total to give the output (see 'Multiple Inputs' below for more on the bias). It is the nonlinear transfer function which allows the network to incorporate nonlinearity. If the transfer function is linear then the network is also linear. This can be used to gauge the improvement in classification gained by allowing nonlinearity. Single neurons are shown in figure 3.1.

{\bf Note:} the equations in this section \emph{without} the ij subscripts are modified from those given in the M{\sc{atlab}} site. The figures in this section are also from the site, modified so that the equations are given separately. Figures 3.5 and 3.6 are slightly modified. The URL for the site is {\tt{http://www.mathworks.com\\/access/helpdesk/help/toolbox/nnet/nnet.shtml}}.

\begin{figure}[!htbp]
\begin{center}
\includegraphics[angle=270, width=12cm]{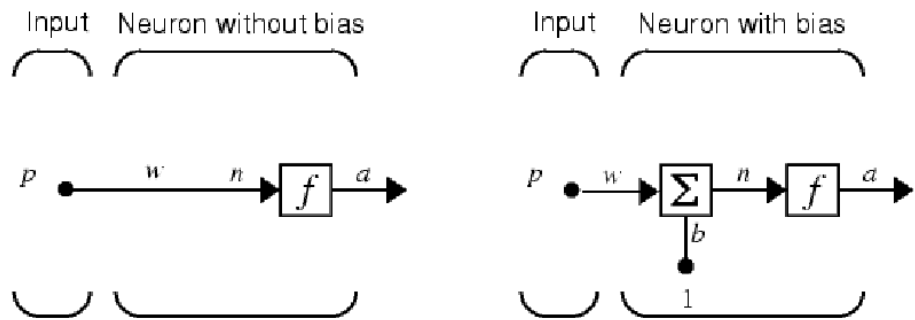}
\caption[Single neurons without and with bias]{Single neurons without and with bias. The $\Sigma$ represents the addition of $wp$ and $b$. In the figures, $n=wp$ and $wp+b$ respectively.}
\end{center}
\end{figure}

For the single neurons \\

\begin{center} $a=f(wp) \qquad$ and $\qquad a=f(wp+b)$ \end{center}

\noindent where \\

\begin{center}
\noindent $a = \textnormal{output}$ \\
$f = \textnormal{transfer function}$ \\
$w = \textnormal{input weight}$ \\
$p = \textnormal{input parameter}$ \\
$b = \textnormal{bias}$ \\
\end{center}

\noindent Examples of transfer functions used in classifying galaxies are:

\begin{itemize}
\item {\bf{Hard limit:}} the output is 0 if the input is below a certain value, and 1 if the input equals or is above the value
\item {\bf{Linear:}} the output is the same as the input, or a linearly scaled multiple of the input
\item {\bf{(Log) sigmoid:}} the output is between 0 and 1, scaling an input in the range $\pm \infty$ by $\frac{1}{1+ e^{-n}}$
\item {\bf{tanh:}} similar to sigmoid, given by $\frac{2}{1+e^{-2n} -1}$. It scales the output to between -1 and 1.
\end{itemize}

In M{\sc{atlab}} (see Chapter 4), these are called {\bf{hardlim}}, {\bf{purelin}}, {\bf{logsig}} and {\bf{tansig}} respectively. The functions are shown in figures 3.2--3.4.

\begin{figure}[!htbp]
\begin{center}
\includegraphics[angle=270, width=4cm]{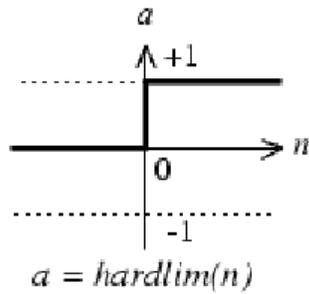}
\caption[The hard limit neuron transfer function]{The hard limit neuron transfer function. Again $n=wp {\textnormal{ or }} wp+b$.}
\end{center}
\end{figure}

\begin{figure}[!htbp]
\begin{center}
\includegraphics[angle=270, width=4cm]{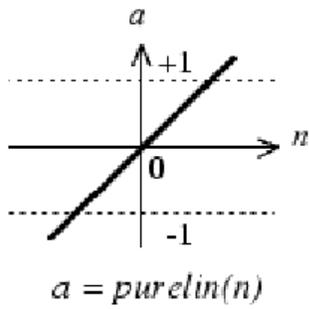}
\caption{The linear transfer function} 
\end{center}
\end{figure}

\begin{figure}[!htbp]
\begin{center}
\includegraphics[angle=270, width=4cm]{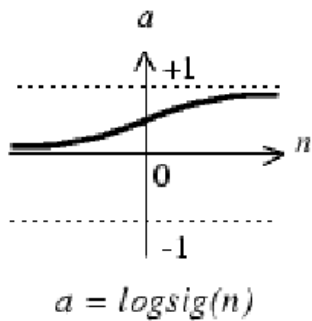}
\caption{The sigmoid transfer function} 
\end{center}
\end{figure}

Clearly many other function are possible, although it has been shown that 'networks with biases, a sigmoid layer, and a linear output layer are capable of approximating any function with a finite number of discontinuities' (M{\sc{atlab}} site). This approximation can be made arbitrarily well. Sigmoids are also differentiable, which is required for the learning algorithm in multilayer networks (see \S 3.6.2).

\subsection{Multiple Inputs}

A multiple input neuron is shown in figure 3.5.

\begin{figure}[!htbp]
\begin{center}
\includegraphics[angle=270, width=6cm]{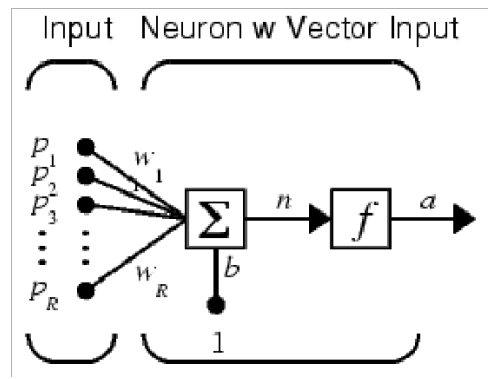}
\caption[A multiple input neuron]{A multiple input neuron ($w$ terms modified from M{\sc{atlab}} site).}
\end{center}
\end{figure}

\noindent Here

\begin{displaymath} a= f({\bf{Wp}}+b) \end{displaymath}

\noindent or

\begin{displaymath} a= f[(\sum_{i=1}^R w_i p_i) + b] \end{displaymath}

Multiple input neurons are similar to a single input neuron, the only difference being that there are $R$ inputs. Each set of $R$ inputs corresponds to one galaxy, and represents a vector in $R$-dimensional space. There is a weight for each input, giving $w_1$, $w_2$, \ldots $w_R$, thus the weights also represent an $R$ dimensional vector. The $wp$ above becomes the dot product ${\bf{Wp}}$ ($w_1 p_1 + w_2 p_2 + ... w_R p_R$). The bias, still a scalar, is added to this product as above. If the data points are plotted in R dimensional space, then the bias allows the line/plane etc. representing the boundary between two classes (the {\bf{decision boundary}}, figure 3.6) not to be forced through the origin. For example, $y=x$ has zero bias, $y=x-1$ has a bias of -1. In two dimensions the function is shifted left by $b$ along the $x$ axis. The vector represented by the $R$ weights is orthogonal to the decision boundary. Thus the weights define the decision boundary, and hence which class each point belongs to. The number of points exactly on the decision boundary should be negligible in the case of classifying galaxies, and even if some were present, any error from this is small compared with the intrinsic spread in the parameters and the classifications. (Galaxies are complex!) A two dimensional decision boundary is shown in figure 3.6.

\begin{figure}[!htbp]
\begin{center}
\includegraphics[angle=270, width=12cm]{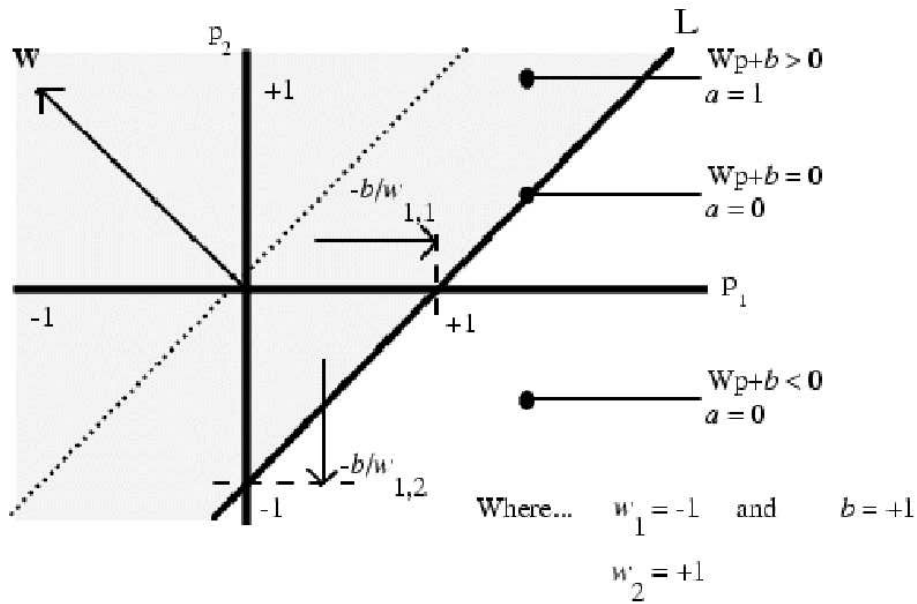}
\caption[A decision boundary in two dimensions]{A decision boundary in two dimensions (w terms modified from M{\sc{atlab}} site).}
\end{center}
\end{figure}

\subsection{Multiple Outputs}

A network may also have multiple outputs ($a_1$, $a_2$ \ldots $a_m$ for $m$ outputs), and a classification is not restricted to being one output. If more than one output is used then it can be shown (e.g. Gish 1990) that for an ideal network (i.e. one which always classifies correctly) the figures generated for each class correspond to Bayesian {\it{a posteriori}} probabilities. An a posteriori probability for an output class is the probability that the output is of that class given that the input parameters are the values they are. Hopefully one class will have a much higher probability than the others, but sometimes two classes may have high values, indicating a galaxy of some intermediate class. For example if the probabilities were, say, \~{}45\% for Sa and \~{}45\% for Sc then the true classification might be Sb (the other \~{}10\% would be the low probabilities for the other classes). This can be useful for checking if classifications are uncertain. A diagnostic for multiple outputs is that the probabilities should add to one for each galaxy. Multiple outputs were not tried in this project. An example is Storrie-Lombardi et al. 1992 (see Chapter 4).

\section{A Layer of Neurons}

The single neuron is easily generalised to several neurons, each with several inputs and outputs. $p$ is a scalar for one input, and an $R$ dimensional vector for $R$ inputs. ${\bf{W}}$ becomes an $S$ by $R$ matrix ${\mathsf{W}}$, for the $S$ neurons in the layer. The bias becomes an $S$ dimensional vector. There are $S$ outputs, so the output is also an $S$-dimensional vector. See figure 3.7.

\begin{figure}[!htbp]
\begin{center}
\includegraphics[angle=270, width=6cm]{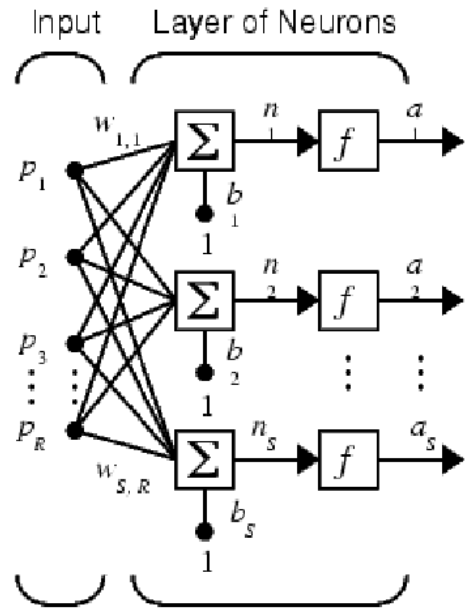}
\caption[A single layer of neurons]{A single layer of neurons (M{\sc{atlab}} site).}
\end{center}
\end{figure}

\noindent For the layer

\begin{displaymath} {\bf{a}} = {\bf{f}}({\mathsf{W}}{\bf{p}} + {\bf{b}}) \end{displaymath}

\noindent which can also be written as

\begin{displaymath} a_k = f_k (\sum_{i=1}^S \sum_{j=1}^R w_{ij} p_j + b_i) {\textnormal{ where }} k=1 \ldots S \end{displaymath}


\section{Multiple Layers of Neurons}

The single layer can then be generalised to multiple layers. The outputs from the neurons in one layer become the inputs for those in the next. There can be a different number of neurons in each layer, and one or more inputs to the first layer and outputs from the last layer. There is a separate weight matrix between each layer, of dimensions $R^l$ by $S^l$ for $R$ inputs from the previous layer and $S$ neurons in the current layer, for layer $l$. The matrix between the inputs and the first layer is the input weight matrix ${\mathsf{IW}}$, and those between layers are the layer weight matrices ${\mathsf{LW}}$. Figure 3.8 shows the arrangement for three layers.

\begin{figure}[!htbp]
\begin{center}
\includegraphics[angle=180, width=12cm]{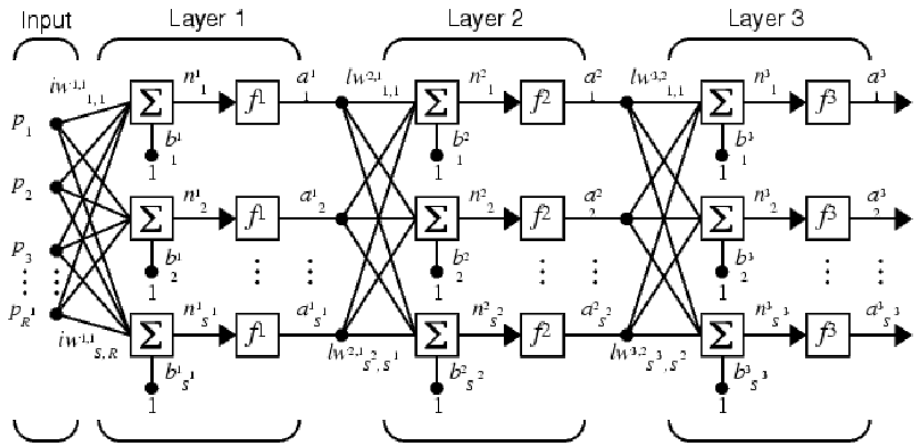}
\caption[Three layers of neurons]{Three layers of neurons. Other numbers of layers are similar. Each layer can have any number of neurons, including one. The superscripts indicate the layer number (1, 2 or 3 here).}
\end{center}
\end{figure}

\noindent For the example of three layers

\begin{displaymath} {\bf{a}}^1={\bf{f}}^1({\mathsf{IW}}^{1,1}{\bf{p}}   +{\bf{b}}^1) \end{displaymath}

\noindent for layer 1, and \\

\begin{displaymath} {\bf{a}}^2={\bf{f}}^2({\mathsf{LW}}^{2,1}{\bf{a}}^1 +{\bf{b}}^2) \qquad \qquad {\bf{a}}^3={\bf{f}}^3({\mathsf{LW}}^{3,2}{\bf{a}}^2 +{\bf{b}}^3) \end{displaymath}

\noindent for layers 2 and 3. \\

Also

\begin{displaymath} {\bf{a}}^3={\bf{f}}^3 ({\mathsf{LW}}^{3,2}{\bf{f}}^2 ({\mathsf{LW}}^{2,1}{\bf{f}}^1 ({\mathsf{IW}}^{1,1}{\bf{p}}+{\bf{b}}^1 )+{\bf{b}}^2 ) +{\bf{b}}^3 ) \end{displaymath}

\noindent where the superscripts refer to the layer of that number, i.e. 1, 2, or 3, and ${\mathsf{LW}^{3,2}}$ represents the layer weight matrix from layer 2 to layer 3, and so on.

These are of similar form for any number of layers, and simplify to the equations earlier for the simpler networks. Multilayer networks are the ones most commonly used for classifying galaxies. The galaxy parameters form the input layer and the classifications the output layer. See Chapter 4 for examples. In this project the output layer is always a single neuron.


\section{Training the Networks}

For a network to classify galaxies correctly, it must perform the correct mapping on the input parameters to produce the output classification. To obtain the correct mapping, or, if the parameters have intrinsic spread, the nearest approximation, the weights and bias must be adjusted to the optimal values. This is either done in an 'unsupervised' manner (see \S 3.7), or, as described here, by {\bf{training}} the network. A set of inputs are given which correspond to a 'correct' set of classifications. This set is called the {\bf{training set}}. The network reads the inputs and produces an output. The weights and bias are adjusted to try to bring them closer to the optimal values, using a chosen {\bf training algorithm}. This process is iterative, and can be summarised by the following

\begin{itemize}
\item Select an appropriate training set
\item Run network on the parameters to produce an output
\item Adjust weights and bias
\item Run network again
\item Continue until weights and bias are near to optimal values
\end{itemize}

\subsection{A Training Set}

The training process causes the network to mimic the typical mappings made in the training set to produce the classifications from the parameters. The training set must therefore be a representative sample of the data, so that the network is able to correctly classify objects which were not used in training. Training takes much more processing time than the subsequent running of a network.

In the case of galaxies, the classification is the galaxy type determined by a human expert and the parameters are various attributes of the galaxy image. An example would be if a galaxy has, say, a low central concentration of light, no discernable bar and quite loosely wound spiral arms then an expert using the broad categories of the Hubble system would call it Sc. Clearly the parameters are more quantitative than this, since the network deals with numerical values. A caution is that the training set should not contain too many idiosyncracies of the particular human classifer, otherwise the network will mimic them. The solution is to average types from more than one person in some quantifiable way.  Typical training sets are a few hundred galaxies, each with of the order of ten parameters. Particular parameters, training sets and the averaging process are described in Chapter 4.

The above assumes that the parameters have been successfully extracted from the galaxy images. Various software packages exist for doing this and in this project only ready-parameterised galaxies are used, both in training and in classification. This is not a serious restriction because sufficient galaxies are available for training. The galaxies in the SDSS Early Data Release are parameterised in this way using the SDSS's dedicated 'Photo' software (Lupton et al. 2001).

\subsection{Training}

The network is created with an initial set of weights and biases, which are, as appropriate, scalars, vectors or matrices of the correct dimension for the network arrangement. They are usually set to zero or to random values. After the first run with the training data the weights and biases must be adjusted to try and move them towards the optimal values for correct classification. This is done using a training algorithm. Typically these involve some kind of {\bf{gradient descent}}, in which the point representing the weights vector is moved down the steepest gradient of the error contours towards the minimum in the error space. An approximate two dimensional analogy is a ball rolling down a sticky hill into a hollow: the ball stops when it reaches the lowest point.

There are many different training algorithms described in the literature, but the one commonly used in galaxy classification is {\bf{backpropagation}}. This tries to minimise the {\bf{cost function}} describing the difference between the classifications output from the network and the true values:

\begin{displaymath} E=\frac{1}{2}\sum_{k}(o_k-d_k)^2 \end{displaymath}

\noindent where $E=\textnormal{error, } o=\textnormal{network output, } d=\textnormal{true output, }$ for a set of $k$ galaxies. The minimisation is done by applying

\begin{displaymath}\delta w_{ij} (t+1)=-\eta \frac{\partial E}{\partial w_{ij}}+\alpha \delta w_{ij} (t) \end{displaymath}

\noindent where $t$ is the iteration number, $\eta$ is the learning rate, $\alpha$ is the 'momentum' of the movement of the error point from the previous iteration, and the $\frac{\partial E}{\partial w_{ij}}$ is the gradient which is being descended. The second term is optional. $\eta$ and $\alpha$ are constants (e.g. Lahav et al. 1996). The corrections are propagated back along the network, hence the term backpropagation.

In this project a more sophisticated variant of the algorithm, known as {\bf resilient backpropagation}, is used. This compensates to some extent for the flatness of the sigmoid transfer function for large or small input values, and in general is faster to converge than simple backpropagation. To the author's knowledge, the algorithm has not been specifically named and used in the galaxy classification literature. Here only simple backpropagation has been used, although other added sophistications such as regularisation (see \S 3.7) have been used in other studies (e.g. Naim et al. (1995b), Lahav et al. (1996)).
The algorithm is available in the M{\sc atlab} Neural Network Toolbox, and is described on the website (PDF documentation chapter 5), and in Riedmiller \& Barun (1993). Another algorithm, the Levenberg-Marquart (Matlab site PDF Chapter 5) was also tried, but was found to be less useful (see \S 4.4.1). Many further algorithms are available on the site, but these two were the ones most recommended by the documentation.

However, no matter which training algorithm is used, it is highly unlikely that any set of weights it finds will be able to map every galaxy onto its correct classification. This is because of the nonlinear nature of the mapping and the intrinsic spread in the properties of galaxies. Therefore a criterion must be used to decide when to stop training the network, otherwise the iteration will just continue indefinitely, with the error tending towards an asymptotic minimum value. Possibilities are

\begin{itemize}
\item Train until the error drops below a certain value
\item Train until the gradient drops below a certain value
\item Train for a fixed number of iterations (or {\bf{epochs}})
\item Train for a fixed amount of processor time
\end{itemize}

In general the minimum which the algorithm tends toward is not necessarily global. It could be a local minimum in the error space which is higher than the global minimum. The gradient descent backpropagation algorithms described here will only descend, so they could get stuck in a local minimum. This can be solved either by using a more sophisticated algorithm, which will be more complex and therefore slower, or by averaging the results of several runs of the network using different random initial weights and bias. It may also be the case that different sets of weights are equally good for classifying. The input parameters can also be normalised to be within a certain range, for example between 0 and 1 (e.g. Naim et al. 1995b). This allows the relative values of the weights to be meaningfully compared, thus giving a method of assessing the relative importance of the various input parameters.

A further point to bear in mind is that the parameters $\eta$ and $\alpha$ are arbitrary, and so good values must be found essentially by trial and error. If $\eta$ is too low the network will not find the optimal weights very quickly and if it is too high the minimum may be overshot, resulting in weights which diverge exponentially from the optimal values.

\section{Further Points}

The networks described above are

\begin{itemize}
\item {\bf{Backpropagation:}} the backpropagation (gradient descent) algorithm, and the more sophisticated variants described, are used for training
\item {\bf{Multilayer:}} the neurons are arranged in layers, with input, hidden and output layers
\item {\bf{Perceptrons:}} the neurons each have a transfer function, either hard limit, linear, sigmoid or tanh
\item {\bf{Supervised:}} the networks are trained with an example set of galaxies classified by humans
\item {\bf{Feedforward:}} the training has no effect on the input parameters, just the weights and biases
\item {\bf{Static:}} there is no time dependence in the network
\end{itemize}

They are thus known as feedforward multilayer perceptrons. (In fact a perceptron neuron originally meant that the neuron had the hard limit transfer function, but a multilayer 'perceptron' can have any set of transfer functions). Clearly many further possibilities exist. Some are now described. More details of principal component analysis, backpropagation, quasi-Newton and the Bayesian perspective are, for example, in the appendices to Lahav et al. 1996. Many of the possibilities are also on the M{\sc{atlab}} site or in Bishop (1995).

\begin{itemize}

\item {\bf{Other networks:}} Many other types of neural network exist, for example radial basis, competitive learning, self-organising maps, learning vector quantisation networks, or time dependent examples such as the Elman and Hopfield networks.

\item {\bf{Transfer functions:}} Besides hard limit, linear, sigmoid and tanh, others include competitive transfer, radial basis, triangular basis, or simply variations such as hard limit between -1 and 1 instead of 0 and 1, negative linear etc. The possibilities are virtually endless.

\item {\bf{Training algorithms:}} Besides the gradient descent and resilient forms of backpropagation, and the Levenberg-Marquardt algorithm, many more routines exist. Examples are the Quasi-Newton algorithms, versions of Newton-Raphson iteration for finding the roots of a quadratic equation, conjugate gradient descent, where the steepest gradient is not necessarily followed, and variable learning rate, which can escape from local minima. None that have been tried have significantly improved the classifications from the networks, although the {\bf{unsupervised}} algorithm is important.

\item {\bf{Unsupervised networks:}} Instead of being trained using subjectively classified data, these types of network look on their own for clustering patterns within the data. They can thus perform classification in a completely objective manner, and may find patterns that the humans have missed. One sort acts as a nonlinear generalisation of the well-known statistical tool of principle component analysis (PCA), in which the linear combination of input parameters with maximum variance is found. Another is the Kohonen Self Organising Map (KSOM). Both PCA and KSOMs have been applied to classifying galaxies at moderate redshifts to try and extend the Hubble sequence in a quantitative way into lookback times where galaxy evolution has become significant (e.g. Abraham et al. 1994, Naim et al. 1997). It has been found that the Hubble sequence breaks down at moderate to high redshift. PCA is important in classifying galaxies spectrally, where the parameters correspond to those describing galaxy evolution, such as star formation rate.

\item {\bf{Regularisation:}} This adds to the cost function to stop the learning rate becoming too high. It is used by Naim et al. (1995b) and others. It is most useful when only small datasets are avilable for training.

\item {\bf{Methods of initialisation:}} Besides intialising the weights and bias to zero, or randomly, algorithms exist to try and optimise the initialisation in some way. An example is the Nguyen-Widrow algorithm.

\item {\bf{Feedback, Time Dependence and Incremental Training:}} These are of less use for galaxies, but have found use in other areas of astronomy, such as predicting solar activity, as mentioned above. In incremental training the weights and biases would be adjusted by training for several epochs on each individual galaxy in turn as it was presented. Thus the network would only remember the galaxy it had just been trained on. The method used above, i.e. adjust after being presented with all the galaxies, is known as batch training. This is clearly preferable.

\item {\bf{Multiple Networks:}} The output from one neural network can form the input for another, or several networks can be used in parallel in ways which are not equivalent to simply having a larger network. Examples are the 'waterfall' arrangement of Adams \& Woolley (1994) and the ensembles of Bazell \& Aha (2001). The latter have been found of give a slight improvement over single networks. The individual networks are given random training sets, and, whilst each one on its own performs less well, the voting system between their outputs results in the correct output being chosen more often. A network is generally more likely to choose the correct output than any particular incorrect one.

\item {\bf{Bayesian Perspective:}} A final point is that the way neural networks have been described in this chapter, and the expressions given, are not the only way they can be described. Equivalent descriptions can be given in terms of probability theory, i.e. a Bayesian perspective. The one example included is that of using multiple outputs instead of one, in which the outputs correspond to a posteriori probabilities. 

\end{itemize}

For classifying galaxies, many of these possibilities have either not been tried or have not been found to be significantly better than those described above. Results from the literature and those from this project are described in Chapter 4.


\chapter[Results]{Results: Humans and Linear Classifiers versus ANNs}

The project aims to try out various artificial neural network programs and see how they compare to each other, to linear classifiers, and to humans in correctly classifying galaxies. This section presents and compares the programs used, the results obtained by previous studies in the literature, and the results obtained by this project.

\section{Programs}

Trying every neural network program from the hundreds available clearly was not feasible since it takes too long to learn to use each one, so emphasis was placed on trying less programs but in more depth. There is no widely used ANN program for classifying galaxies,  so an early task of the project was to see what was available on the web. Another constraint is that it would take too long to learn and run image analysis software, so ready-parameterised images were used. In fact, it turned out that only \emph{one} program was required, because it allowed the generic creation of almost any neural network. This program is the M{\sc{atlab}} Neural Network Toolbox, described below.

\subsection{LMorpho}

LMorpho, or Linux Morpho, was the first program tried. It is a collection of routines for working with and classifying galaxy images written by S.C. Odewahn at Arizona State University, USA, and includes a neural network routine. It is available for download at {\tt{http://www.public.asu.edu/\~{}asu-\\sco/documents/lmorpho/dist/index.html}}. Unfortunately, the program was described as 'not particularly user friendly' by the author, and some time was spent on installation, configuring the compiler, and getting the program to work. It then turned out that full image processing would be required to extract parameters in a suitable form to run through the network. This, and the availability of the M{\sc{atlab}} program described below meant that this program was abandoned for this project. The version tried was lmorpho\_jul24\_2000, run on Red Hat Linux 6.2.

\subsection{M{\sc{atlab}} Neural Network Toolbox}

This set of tools within the well known M{\sc{atlab}} program enabled the generation from scratch of many types of neural network, including those useful for galaxy classification. Extensive documentation is available on line, including much introductory material. Results are detailed in \S 4.4. The version used was 5.3.1.29215a (R11.) from Oct 6th 1999. The URL for the documentation is {\tt{http://www.mathworks.com/access/helpdesk/help/toolbox/nnet/\\nnet.shtml}}.

\subsection{Other Programs}

The range of options available in and the generic-ness of the networks in M{\sc atlab} meant that no other programs needed to be tried out. Others do exist (see \S 5.1).

\section{Galaxy Parameters}

The number of parameters used to describe a galaxy can vary from one to $\infty$. Some kind of compromise must therefore be found which retains most of the physical information. Many different parameters are used, but usually of order one to ten per galaxy.

If an inappropriate set is used information is lost and the networks cannot classify properly. If too many are used the networks are slow, and in training are more likely to get stuck in local error minima and mimic noise in the data, both of which render them unable to classify properly new galaxies which they have not previously seen.

In this project the parameters are standard outputs from the SDSS photometry software {\bf{Photo}} v5\_0\_3 as used in Shimasaku et al. (2001) (given by them as 'late 1999', see also Lupton et al. (2001)). These include the $r_{50}/r_{90}$ (inverse) concentration index, likelihood of de Vaucouleurs profile $P_{DeV}$, likelihood of exponential profile $P_{Exp}$, and colour in the five SDSS bands $u^{*}, g^{*}, r^{*}, i^{*}, z^{*}$. These parameters are now described. \\

The {\bf{concentration index}} defined in Shimasaku et al. (2001) is based on the {\bf{Petrosian radius}} $r_{P}$. This is independent of galaxy distance, because it is based on galaxy surface brightness, and is given by the implicit expression

\begin{displaymath} \eta =\frac{I(r_{p})}{2\pi \int_{0}^{r_{P}}I(r)rdr/(\pi r^2_{P})} \end{displaymath}

\noindent where $r=\textnormal{linear radius}$, $I=\textnormal{intensity}$, $\eta={\textbf{Petrosian Ratio}}$ (see below).

This is for an annulus of infinitesimal width and finite diameter within the object. For a value of $r_{P}$ in practice, values are adopted such that the local surface brightness in an annulus of finite width from $\alpha r_{P}$ to $\beta r_{P}$ is $\eta$ percent of the mean surface brightness within $r_{P}$. The mean surface brightness is the {\bf{Petrosian flux}} $f_{P}$

\begin{displaymath} f_{P}=2\pi\int_{0}^{2r_{P}}I(r)rdr \end{displaymath}

\noindent per unit area $\pi r^2$, i.e.

\begin{displaymath} 2\pi\int_{0}^{2r_{P}}I(r)rdr/(\pi r^2) \end{displaymath}

\noindent thus

\begin{displaymath} \eta(r)=\frac{2\pi\int_{\alpha r}^{\beta r}I(r)rdr/[\pi(\alpha^2-\beta^2)r_{P}^2]}{2\pi\int_{0}^{r_{P}}I(r)rdr/(\pi r_{P}^2)} \end{displaymath}

\noindent Usually $\eta=0.2$, $\alpha=0.8$, $\beta=1.25$. These values are used in the literature, including Shimasaku et al., and are hence used here. \\

The concentration index is then given by the ratio of the radii $r_a /r_b$ where the radii are those at which $a$ and $b$ percent of the total Petrosian flux are received. Usually $a=50$ and $b=90$, giving $r_{50}/r_{90}$. \\

\noindent (Petrosian equations adapted from Shimasaku et al. (2001) and Strateva et al. (2001).) \\

The {\bf{de Vaucouleurs profile}} (de Vaucouleurs 1948) describes the radial distribution of intensity in elliptical galaxies, and is given by:

\begin{displaymath} I(r)=I_0 \textnormal{exp} \{ -7.67[(r/r_e )^{1/4}-1] \} \end{displaymath}

\noindent where $I(r)$=intensity at radius $r$, $I_0$ = central intensity, $r_e$= a the half-light radius for the galaxy. \\

The {\bf{exponential profile}} (e.g. Freeman 1970) is similar, but for spiral galaxy disks:

\begin{displaymath} I(r)=I_0 \textnormal{exp}(-1.68r/r_e) \end{displaymath}

So $P_{DeV}$ and $P_{Exp}$ correlate with whether a galaxy is spiral or elliptical. The probabilities are standard $\chi ^2$ fits to the respective profiles. \\

{\bf{Colours}} are given as differences between magnitudes in defined wavelength bands for a galaxy. Shimasaku et al. (2001) use the preliminary Sloan Digital Sky Survey (SDSS) bands $u^{*}$, $g^{*}$, $r^{*}$, $i^{*}$ and $z^{*}$. These are shown in table 4.1. The values used in the project are not corrected for galactic extinction, and the mean reddenings in the SDSS Early Data Release (EDR) for each band are 0.21, 0.15, 0.11, 0.08, and 0.06 magnitudes. However, this is small compared with the intrinsic dispersions in the colours and in the context of using neural networks. Also, the preliminary SDSS magnitudes are so-called because the calibration is preliminary. The final magnitudes may differ by as much as 0.1.

\begin{table}[!htbp]
\begin{center}
\caption[The SDSS Photometric System]{The SDSS Photometric System (Stoughton et al. (2001), table 20). FWHM = full-width half-maximum, a measure of the bandwidth.}
\gap
\begin{tabular}{|c|c|c|}
\hline
Band &Wavelength/{\small\AA} &FWHM/{\small\AA} \\
\hline \hline
$u^*$ &3551 &581 \\
$g^*$ &4686 &1262 \\
$r^*$ &6166 &1149 \\
$i^*$ &7480 &1237 \\
$z^*$ &8932 &994 \\
\hline
\end{tabular}
\end{center}
\end{table}

The independent colours are therefore $u^*-g^*$, $g^*-r^*$, $r^*-i^*$ and $i^*-z^*$. 

\noindent Many more parameters have been used in the literature, for example

\begin{itemize}
\item {\bf{Isophotes:}} various values of Petrosian radii could be measured, e.g. $r_{10}$, $r_{20}$ \ldots $r_{90}$, or $\gamma$ in the concentration parameter expression above could be varied.
\item {\bf{Absolute magnitudes:}} or Petrosian fluxes could be given in a similar way to the radii.
\item {\bf{Spiral arm pitch angle:}} The angle between the spiral arm and the tangential direction to the galaxy centre at the point where the arm starts can be quantified.
\item {\bf{Harmonics:}} Block et al. (1999) use harmonics in spiral galaxy disks, viewed in the infrared so that the dust extinction is low, as a basis for classifying spiral disks.
\item {\bf{Bar strength:}} An example is Abraham \& Merrifield (2000), who define the bar strength as $\frac{2}{\pi}[arctan(b/a)^{-1/2}_{bar} - arctan(a/b)^{-1/2}_{bar}]$. This gives zero for no bar and one for an infinitely strong bar.
\item {\bf{Measures of asymmetry:}} The image of a galaxy is projected on the sky, so measurements such as the ratio of the longest diameter to the shortest for an image can quantify how face-on or edge-on the image is. Edge on disk galaxies are difficult or impossible to classify.
\item {\bf{Octants:}} The profile fits, etc. can be measured for octants or quarters, etc. of the galaxy image. Storrie-Lombardi et al. (1992) fit a generalised de Vaucouleurs profiles to various numbers of octants (the 'oct' parameters in figure 4.1).
\end{itemize}

The parameters assume that effects such as extinction and reddening of light by intervening dust have been accounted for. Maps exist of galactic extinction (e.g. Schlegel et al. 1998), although the correction is still a potential source of error. Note again that the colours used in \S 4.4 have not been corrected, but the effect is small in this context.

\section{Results from Previous Studies}

Various quantitative classification studies using automated classification techniques have been carried out since the early 1980s. An important example, and one of the first to use neural networks, is Storrie-Lombardi et al. (1992). They used a feedforward backpropagation network with various configurations, including 13:13:5 (13 inputs, 13 neurons in first layer and 5 outputs), shown in figure 4.1. They compared the outputs from this network with the classifications of human experts for 5217 galaxies from the European Southern Observatory catalogue (ESO-LV, Lauberts \& Valentijn 1989). They found that the network agreed with the humans 64\% of the time and agreed to within one class (of the 5) 90\% of the time (table 4.2). The ESO-LV classifications were given in the catalogue itself and so were separate from the Storrie-Lombardi et al. project. The 64\% compares with 56\% from the ESO-AUTO automatic classifier ESO-AUTO, which used conventional linear statistics as opposed to an ANN. Thus the ANNs showed a significant improvement.

\begin{figure}[!htbp]
\begin{center}
\includegraphics[angle=90, width=12cm]{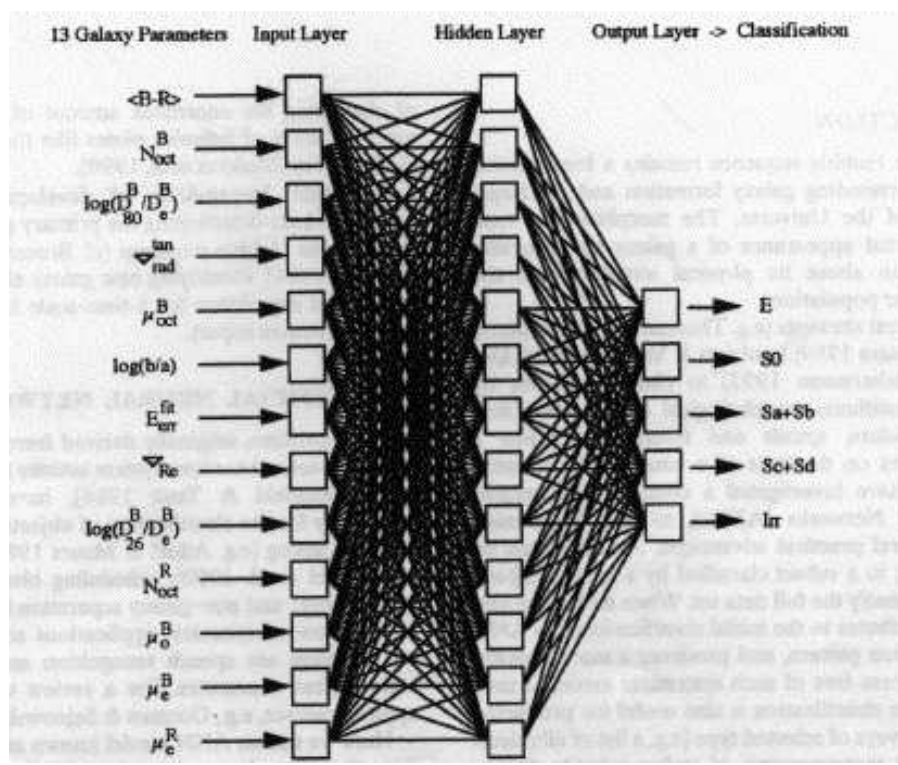}
\caption[13:13:5 network used by Storrie-Lombardi et al. (1992)]{A network used by Storrie-Lombardi et al. (1992). This one is 13:13:5. Others are similar. (From Storrie-Lombardi et al. 1992.)}
\end{center}
\end{figure}

\begin{table}[!htbp]
\begin{center}
\caption[Humans (ESO-LV), versus ANN and ESO-AUTO, from Storrie-Lombardi et al. 1992]{Humans (ESO-LV) in rows, versus ANN and ESO-AUTO in columns, modified from Storrie-Lombardi et al. 1992. The diagonals give the classifications which agree, and amount to 64\% of the total for the ANNs and 56\% for ESO-AUTO. For classifications within one class, the agreement is 90\%.}
\gap
{\small
\begin{tabular}{|c||c|c|c|c|c||c|c|c|c|c|}
\hline
&\multicolumn{5}{c||}{Humans vs. ANNs} &\multicolumn{5}{c|}{Humans vs. ESO-AUTO} \\
\hline \hline
Class & E & S0 & Sa+Sb & Sc+Sd & Irr   & E   & S0  & Sa+Sb & Sc+Sd & Irr         \\
\hline
E     & 203 & 77  & 25   & 1   & 5     & 197 & 87  & 17    & 5     & 5           \\
S0    & 109 & 229 & 240  & 7   & 2     & 184 & 218 & 155   & 28    & 2           \\
Sa+Sb & 12  & 85  & 1281 & 218 & 15    & 106 & 12  & 791   & 664   & 38          \\
Sc+Sd & 1   & 4   & 304  & 415 & 36    & 22  & 11  & 24    & 631   & 72          \\ 
Irr   & 0   & 0   & 53   & 69  & 126   & 22  & 9   & 31    & 42    & 144         \\ 
\hline
\end{tabular}
}
\end{center}
\end{table}

The ANNs also followed the distribution of types much better than the linear ESO-AUTO, again demonstrating their increased capacity to classify comparably with humans (figure 4.2).

\begin{figure}[!htbp]
\begin{center}
\includegraphics[angle=270, width=12cm]{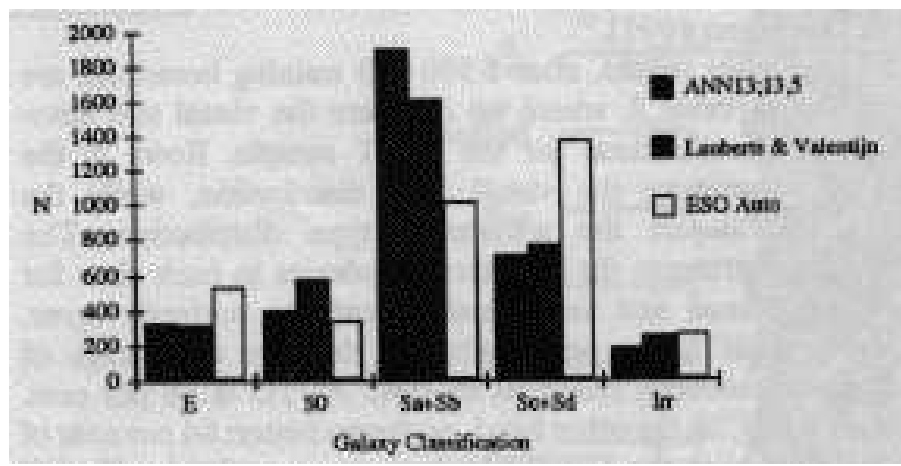}
\caption[Number of galaxies versus type, for the 13:13:5 ANN, ESO-LV and the linear ESO-AUTO (Storrie-Lombardi et al. 1992)]{Number of galaxies versus type, for the 13:13:5 ANN, ESO-LV and the linear ESO-AUTO (Storrie-Lombardi et al. 1992). Note that the origninal figure is also in grayscale.}
\end{center}
\end{figure}

No human trials were carried out in this project, but a trial has been carried out by Naim et al. (1995a). This study created a uniform sample of 831 nearby bright galaxies from photographic plates from the APM galaxy survey (Maddox et al. 1990). Laser prints of the galaxy images were given to six human experts and they were asked to classify the galaxies according to the de Vaucouleurs T system (described in \S 2.3.1). (In fact one of the experts (van den Bergh) used computer screen images and the DDO system, later converted to T-type, but the difference is negligible). It was found that there was on average a root mean square (RMS) deviation of 1.8 T types between the experts.

The mean square deviation (variance) between two observers used is given by \begin{displaymath}\sigma^{2}_{ij}=\frac{1}{N_{gal}}\times\sum_{gal}(T_{i}-T_{j})^{2}\end{displaymath}

\noindent where $N_{gal}$ is the number of galaxies classified by both observers. The RMS values between the experts in Naim et al. (1995b) are shown in table 4.3. The higher values between the experts and the RC3 classifications of the same galaxies illustrates the usefulness of a uniform sample (the RC3 uses a different image set). Subjective classifications will depend on the appearance of the image to some degree. (RC3 = the Third Reference Catalogue of Bright Galaxies, de Vaucouleurs et al. 1991).

Naim et al. (1995b) then studied the classifications produced by various ANNs. The comparison is also described in Lahav et al. (1995). This study was the first and only systematic comparison based on a uniform sample of galaxy images, and presented to several experts from different 'schools of thought'. The results are in table 4.4.

\begin{table}[!htbp]
\begin{center}
\caption[Root mean square deviations between experts' classifications, in Naim et al. (1995b)]{Root mean square deviations between experts' classifications and between the experts and the RC3 catalogue (Lahav et al. 1995, table 2), RB = R. Buta; HC = H. Corwin; GV = G. de Vaucouleurs; AD = A. Dressler; JH = J. Huchra; RC3 = 3rd Reference Catalogue of Bright Galaxies (de Vaucouleurs et al. 1991)}
\gap
\begin{tabular}{|c|c|c|c|c|c|c|}
\hline
Classifier & RB & HC & GV & AD & JH & vdB      \\
\hline \hline
RC3        & 2.2 & 2.1 & 1.8 & 2.3 & 2.2 & 2.4 \\
RB         &     & 1.3 & 1.6 & 1.7 & 1.8 & 1.7 \\
HC         &     &     & 1.5 & 1.8 & 1.9 & 1.9 \\
GV         &     &     &     & 1.7 & 1.8 & 1.9 \\
AD         &     &     &     &     & 2.1 & 1.8 \\
JH         &     &     &     &     &     & 2.0 \\
\hline
\end{tabular}
\end{center}
\end{table}

\begin{table}[!htbp]
\begin{center}
\caption[RMS deviations between the experts and the ANN (Naim et al. 1995b)]{RMS deviations between the experts and the ANN. The ANN is a 13:5:1 configuration with 10 runs averaged with different initial weights. Other configs. e.g. 13:13:1 gave similar results (Lahav et al. 1995 table 3). 'Mean' = when the ANN is trained and run on the mean of the human types, with 'a few outliers removed.' It does not equal the mean of the other columns in the table.}
\gap
\begin{tabular}{|c|ccccccc|}
\hline
                 & RB  & HC  & GV  & AD  & JH  & vdB & Mean \\
\hline \hline
ANN              & 1.9 & 2.0 & 2.2 & 1.9 & 2.3 & 2.2 & 1.8  \\
$N_{gal}$ of 831 & 764 & 812 & 473 & 814 & 824 & 549 & 831  \\
\hline
\end{tabular}
\end{center}
\end{table}

This study finds that 'on the whole there is a reasonable consistency in the way people classify galaxies, but the scatter is significant' and that 'the ANNs can replicate the expert's classification of the APM sample as well as other colleagues or students of the expert'. This suggests that the network was about as capable as the humans of classifying galaxies correctly using the T type system. Other statistics are also given, including a crude estimate of the internal scatter or reproducibility of the classifications of an indivdual observer using a dataset with a lower resolution. One can also plot the network versus itself with a different set of random weights, and perform various other tests.

Many other studies have been made in which galaxies are morphologically classified by some kind of automatic system, including neural networks (e.g. Goebel et al. (1989), Thonnat (1989), Spiekermann et al. (1992), Doi et al. (1993), Serra-Ricart et al. (1993), Abraham et al. (1994), Adams \& Woolley (1994), Odewahn et al. (1996), Naim et al. (1997), Bazell \& Aha (2001)), and many more still have looked at ways to quantify galaxy properties, but Naim et al. (1995b)/Lahav et al. (1995) is the only systematic comparison between ANNs and a large number of independent experts.

\section{Project Results and Discussion}

\subsection{Networks \& Shimasaku Types}

Shimasaku et al. (2001) have obtained eyeball classifications for 456 galaxies from the Sloan Digital Sky Survey (SDSS) by having four of their experts compare the images with those of the Frei and Gunn galaxy catalogue (Frei et al. 1996), which are given RC3 types. In this project these eyeball types are compared with the M{\sc{atlab}} network results which used the parameters for the images (parameters to be published in Fukugita et al., in preparation). The Shimasaku et al. classifications are coarser than the T system but they consider them 'sufficient for most purposes for galaxy science'. They are designated 'Shimasaku Types' here, and are shown in table 4.5.

\begin{table}[!htbp]
\begin{center}
\caption[T-type versus type from Shimasaku et al. (2001)]{T-type versus type from Shimasaku et al. (2001). The name 'Shimasaku Type' is adopted here for their types.}
\gap
\begin{tabular}{|c|cccccccccc|}
\hline
Hubble Type    &   &E0 &   &   &S0  &S0 &    &Sa &   &Sb    \\
\hline
T type         &-6 &-5 &-4 &-3 &-2  &-1 & 0  &1  &2  &3     \\
\hline
Shimasaku Type &   &0  &   &   &1   &1  &    &2  &   &3     \\ 
\hline \hline
Hubble Type    &   &Sc &   &   &Sdm &   & Im &   &   &      \\
\hline
T type         &4  &5  &6  &7  &8   &9  & 10 &11 &90 &99    \\
\hline
Shimasaku Type &   &4  &   &   &5   &   &  6 &   &   &99.99 \\
\hline
\end{tabular}
\end{center}
\end{table}

Two galaxies, which had one or more Petrosian magnitudes of '99.99', were removed to leave a set of 454. (The 99.99 is so far above the other values of 15-20 that when these parameters were run on their own, as below, the 99's seriously affected the results). Several other 'outliers' which had magnitudes of around 22 in the wrong magnitude column were left in, since most real data is likely to have outliers of this sort, due to imperfect photometry, and the networks could 'cope' with these.

The parameters were initially run through a single linear neuron ('purelin' transfer function). This acts as a linear classifier and thus provides a 'control set' against which the improvements achieved by the networks can be seen. The parameters used are given in table 4.6. Note that a single apparent magnitude, r*, is included. This is not expected to correlate with galaxy type but may help by, for example, causing the network to give greater weight to bright galaxies, in which the parameters are better resolved, and thus perhaps more likely to show consistent patterns. The $r^*$ is used because the Petrosian magnitudes in all five bands are defined using $r^*$ so that all the bands can be observed using the same aperture.

Typical combinations of parameters and results for the single neuron are given in table 4.7, for training and classification on the 454 galaxies. The results in 4.7 are arranged such that each subsequent set of parameters shown gave an improved result. For linear neurons, the error space has only one minimum, which is global. Hence there is no need to do several runs for different random initial weights. It is merely necessary not to set the learning rate too high (which gives diverging weights), and to train for sufficient time that the value of the minimum becomes clear. In M{\sc atlab} the linear neuron is always trained using the by weight and bias function 'trainwb', and no gradient value is output. Thus the number of epochs given in the table is that at which the fifth decimal place of the mean squared error (MSE, which \emph{is} outputted) does not change for the first time, to the nearest 25 epochs. From the shape of the training curve (figure 4.3), one can see that the value of the global minimum is always clear at this point.

\begin{figure}[!htbp]
\begin{center}
\includegraphics[width=12cm]{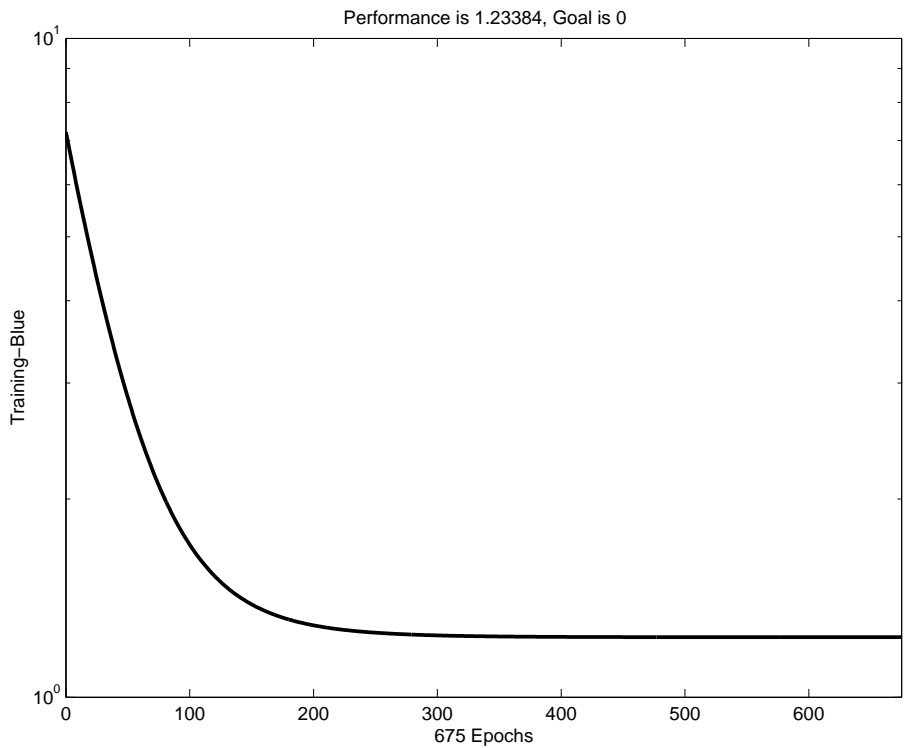}
\caption[A typical ANN training curve]{A typical ANN training curve, showing the mean squared error between the network output galaxy type and the Shimasaku Type, versus number of epochs of training. For a single neuron, as shown here, there is only one minimum reached, which is global. More complex networks have more jagged training curves, but they retain the same form. The neuron here has clearly reached its minimum.}
\end{center}
\end{figure}

From table 4.7, it is clear that the more parameters are used, the better the fit the network is able to make to the training set. However, it is also clear that fairly good fits can be achieved by using just one or two parameters, in particular, the (inverse) concentration index $r_{50}/r_{90}$ (parameter 1) and the $g*-r*$ colour (parameter 5) provided good fits for these 454 galaxies. The importance of the concentration index is in agreement with numerous previous studies. Values in the table are given to 3 significant figures but probably only the first is important. Plots of network type versus true type for the concentration index, and for all 8 parameters are shown as figures 4.4 and 4.5.

\begin{figure}[!htbp]
\begin{center}
\includegraphics[width=12cm]{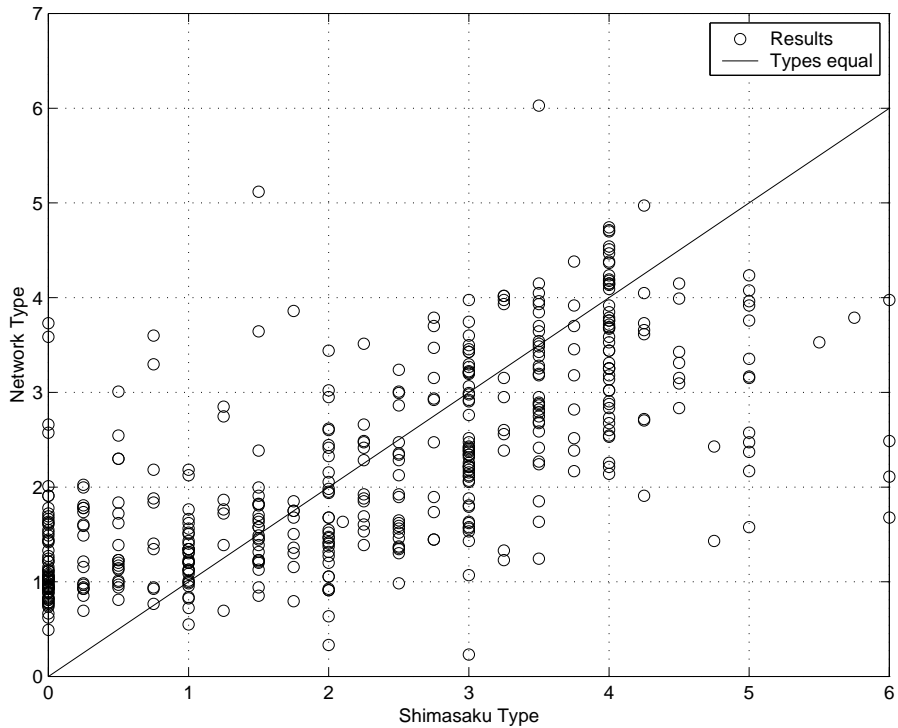}
\caption[Network type versus Shimasaku type for the single linear neuron using just the concentration index as input]{Network type versus Shimasaku type for the single linear neuron using just the concentration index as input. Note the distortion at low and high Shimasaku type values.}
\end{center}
\end{figure}

\begin{figure}[!htbp]
\begin{center}
\includegraphics[width=12cm]{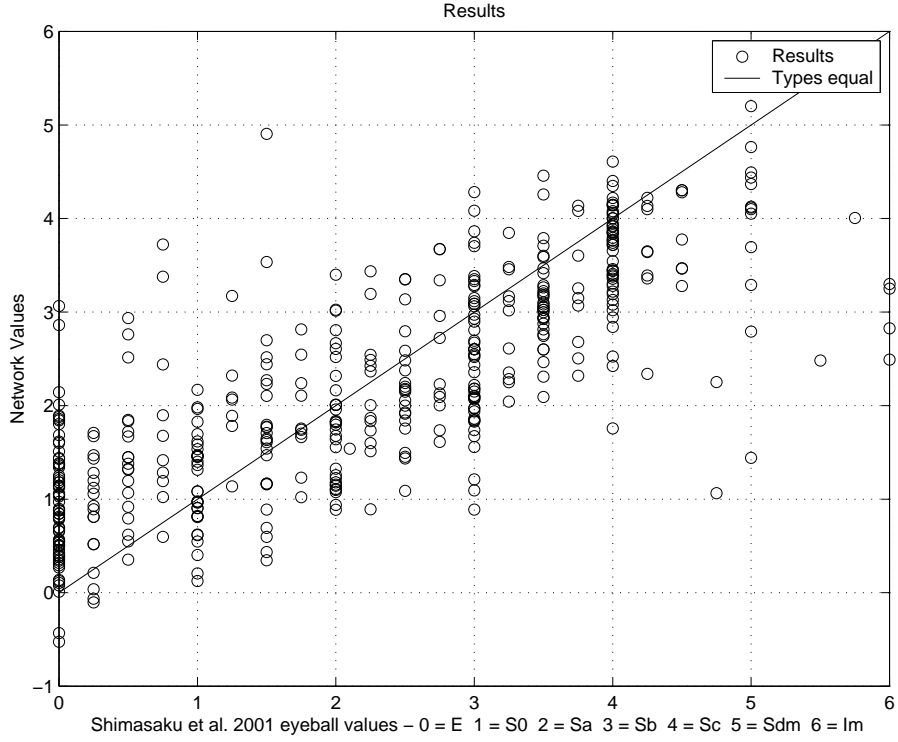}
\caption[Network type versus Shimasaku type for the single linear neuron using all 8 parameters as input]{Network type versus Shimasaku type for the single linear neuron using all 8 parameters as input. Runs from the actual networks in fact look very similar: improved correlation gradually moves the 0 types towards the 'types equal' line, moves most of the 6's up towards the same line, and gives a narrower spread about the line. By the time the correlation looks visually significantly better the network is probably overfitting the training set.}
\end{center}
\end{figure}

The mean squared error (MSE) and root mean square error (RMS) between network type and Shimasaku Type is given for each parameter set. These RMSs between seven types can be compared with the Naim et al. 1995b RMS of 1.8 in 16 types. 1.8 in 16 is equivalent to 0.79 in 7. Thus the linear neuron on its own is not as good, as expected.

\begin{table}
\begin{center}
\caption{Parameters of the networks used in this project}
\gap
\begin{tabular}{|c|c|}
\hline
No &Parameter       \\
\hline \hline
1  &$r_{50}/r_{90}$ \\
2  &$P_{deV}$       \\
3  &$P_{Exp}$       \\
4  &$u^*-g^*$       \\
5  &$g^*-r^*$       \\
6  &$r^*-i^*$       \\
7  &$i^*-z^*$       \\
8  &$r^*$           \\
\hline
\end{tabular}
\end{center}
\end{table}

\begin{table}
\begin{center}
\caption[Results for the single neuron]{Parameters used and results for the single neuron. Compare the RMS column with the equivalent (to 1.8 T types in the 16 from -5 to 10) 0.79 value for that between humans and between humans and ANNs in Naim et al. (1995b). N.B.: MSE (not shown) = mean squared error between network type and Shimasaku type; RMS values are given to 2 d.p.; the number of epochs is that at which the first occurrence of the fifth decimal place of the MSE not changing occured. The RMS values shown are equal to $\sqrt{MSE}$.}
\gap
{\footnotesize
\begin{tabular}{|c|c|c|c|c|}
\hline
Parameter(s)                             &RMS (Shimasaku        &Correlation              &Maximum               &Number of          \\
used                                     &Types)                &Coefficient              &learning rate         &epochs trained     \\
\hline \hline
3                                        &1.51                  &0.238                    &0.0043                &200                \\
7                                        &1.49                  &0.285                    &0.0042                &775                \\
2                                        &1.45                  &0.354                    &0.0042                &175                \\
4                                        &1.44                  &0.373                    &0.0012                &300                \\
2 3                                      &1.38                  &0.451                    &0.0042                &200                \\
6                                        &1.33                  &0.510                    &0.0038                &275                \\
5                                        &1.19                  &0.643                    &0.0029                &225                \\
1                                        &1.11                  &0.698                    &0.0038                &700                \\
1 2 3                                    &1.06                  &0.731                    &0.0037                &775                \\
1 5                                      &1.02                  &0.752                    &0.0027                &1275               \\
1 5 6                                    &1.01                  &0.751                    &0.0024                &1450               \\
1 2 3 5                                  &0.99                  &0.769                    &0.0026                &1750               \\
1 2 3 5 6                                &0.98                  &0.776                    &0.0024                &1875               \\
1 2 3 4 5 6                              &0.98                  &0.777                    &0.00099               &4075               \\
1 2 3 4 5 6 7                            &0.97                  &0.778                    &0.00099               &4100               \\
1 2 3 4 5 6 7 8                          &c. 0.95               &c. 0.791                 &0.000017              &after 95300        \\
\hline
\end{tabular}
}
\end{center}
\end{table}

The parameter set 1 2 3 4 5 6 7 8 was then tested with various actual neural networks. Details are given in table 4.8, and results in table 4.9. The results show that the larger networks are able to classify the galaxies essentially arbitrarily well. However, they are almost certainly 'overfitting' the data. That is, the large networks have so many weights that they are able to 'remember' each galaxy individually rather than rely on generalisations. Thus they would be unable to cope when presented with a new set of galaxies. These galaxies would take a different path through the weight space and be classified essentially at random.

The point to note is that the networks (larger and smaller), when only using some of the parameters, can equal the performance of the single neuron when it uses them all, and for the same parameter set improve on the single neuron, e.g. from $0.7 \pm \sim 0.1$ to $0.8 \pm \sim 0.1$ for just $r_{50}/r_{90}$. It is unfortunate that when they significantly improve by using all the parameters they are, with this training set, overfitting the data. The networks are also much quicker to train than the single neuron when using 8 or more parameters, as can be seen in tables 4.7 and 4.9.  

There was found to be very little difference between the performance using sigmoid and using tanh transfer functions. Sigmoids were chosen because they map positive inputs onto an output between 0 to 1, and with the exception of a few 'outlier' colours, the galaxy parameters are all positive. Because normalisation was not carried out, the outputs were always kept as pure linear so that they could range from 0 to 6 (the Shimasaku types), as opposed to the 0 to 1 of a sigmoid. The Levenberg-Marquardt algorithm converged more quickly in small networks, but was unsuitable for large networks and sometimes failed. One can also see from table 4.9 that adding more neurons in a layer is generally more beneficial than adding more layers. The training times are given. The time taken to classify the 454 galaxies varies from about 8 to 12 seconds depending on the complexity of the network.

Ideally the means should be from a larger number of runs than five, but the standard deviations are shown in the table for the MSE's and correlations, and confirm that the more complex networks do indeed produce a lower MSE. The standard deviations could also be calculated for table 4.7.

A listing from which all the networks shown here can be generated is shown in Appendix B.

\begin{table}
\begin{center}
\caption[Summary of further network details]{Summary of network details. Note: the architectures are S:1 etc. where S = no. of neurons in the layer. The number of inputs and outputs are not shown (always 8 and 1 respectively). Transfer functions: l = sigmoid, p = pure linear, t = tanh; training: lm = Levenberg-Marquardt, rp = resilient backpropagation; those unmarked are 'lp\_rp'. An example: 8:1\_tp\_rp is 8 neurons in the first layer, one in the second (as output), the transfer functions are tansig, purelin and the training algorithm is resilient backpropagation.}
\gap
\begin{tabular}{|c|l|}
\hline
Architectures       &1\_p, 8:1\_tp\_lm, 8:1\_tp\_rp, 8:1\_lp\_rp, S:1 and S:S:1  \\
                    &where S=8, 16, 32 and 64, (except 64:64:1),                 \\
                    &128:128:128:128:1                                           \\
Initial weights     &Here in fact all 1 for p\_1 (better is zero but             \\  
                    &makes no difference to the global minimum),                 \\
                    &randomised by M{\sc atlab} network initialisation           \\
                    &for other architectures                                     \\
Trained on          &454/454                                                     \\
Training algorithms &p\_1: trainwb, rest: trainlm and trainrp                    \\
Learning rate       &p\_1: maximum without being weights diverging,              \\
                    &rest: default                                               \\         
Other parameters    &M{\sc atlab} defaults (see Appendix B)                      \\
\hline
\end{tabular}
\end{center}
\end{table}

\begin{table}
\begin{center}
\caption[Results for various network architectures]{Results for various network architectures when applied to the 454 galaxy training set using all 8 parameters. Each architecture was trained for 1000 epochs. As in table 4.7, compare the RMS (again = $\sqrt{MSE}$) column with the equivalent (to 1.8 T types in the 16 from -5 to 10) 0.79 value for that between humans and between humans and ANNs in Naim et al. (1995b). N.B.: Transfer functions: l = sigmoid (logsig), p = pure linear, t = tanh; training: lm = Levenberg-Marquardt, rp = resilient backpropagation; those unmarked are 'lp rp'. The architecture notation is the same as in table 4.8, where it is explained. The standard deviations are sample standard deviations. Also note: [1] The trainlm failed twice in the 5 runs, so the result here is from 3; [2] the 128:128:128:128:1 was only run once, and simply shows an extreme example of overfitting.}
\gap
{\footnotesize
\begin{tabular}{|c|c|c|c|c|}
\hline
Network           &\multicolumn{2}{c|}{RMS \& Stdev from 5}    &Correlation     &Time to train for \\
                  &\multicolumn{2}{c|}{runs (Shimasaku Types)} &Coefficient     &1000 epochs       \\
\hline \hline
1                 &1.11  &0          &0.698     &0:08      \\
8:1 tp rp         &0.869 &0.0374     &0.828     &0:28      \\
8:1 tp lm         &0.843 &0.0276 [1] &0.836     &1:22      \\
8:1 lp rp         &0.847 &0.0222     &0.844     &0:28      \\ 
16:1              &0.808 &0.0327     &0.857     &0:37      \\
8:8:1             &0.795 &0.0328     &0.858     &0:44      \\
32:1              &0.771 &0.0358     &0.867     &0:55      \\
8:8:8:1           &0.764 &0.0775     &0.869     &0:58      \\
64:1              &0.703 &0.0089     &0.891     &1:21      \\
16:16:1           &0.698 &0.0390     &0.893     &0:58      \\
16:16:16:1        &0.622 &0.0658     &0.915     &1:19      \\
32:32:1           &0.558 &0.0627     &0.932     &1:40      \\
64:64:1           &0.516 &0.0168     &0.943     &3:06      \\
32:32:32:1        &0.444 &0.0164     &0.958     &2:18      \\
128:128:128:128:1 &0.106 &n/a    [2] &0.998     &c. 20 min \\
\hline
\end{tabular}
}
\end{center}
\end{table}

\subsection[Application to the SDSS EDR]{Application to the SDSS Early Data Release}

The networks were then applied to the Sloan Digital Sky Survey Early Data Release (SDSS EDR, Stoughton et al. 2001). A set of galaxies was selected from the catalogue archive server ({\tt http://archive.stsci.edu/sdss/so-\\ftware/)} using the SDSS Query tool and the query \\

\noindent {\tt SELECT ra, dec, z, zErr, zConf, zStatus, tag.petroMag, \\ tag.reddening, tag.petroR50\_r, tag.petroR90\_r, tag.lExp\_r, \\ tag.lDeV\_r, tag.obj.field.segment.run,\\ tag.obj.field.segment.rerun, tag.obj.field.segment.camCol, \\ tag.obj.field.field, tag.obj.objid, tag.rowC, tag.colC \\
FROM specobj \\
WHERE (primTarget == AR\_TARGET\_GALAXY)} \\

This returned a set of 29,433 galaxies with spectra and redshifts. These were chosen a) to provide a reasonably sized dataset and b) so that the distribution of the galaxies could potentially be plotted by the morphological type determined by the networks. Four galaxies were removed from the set, because they had values of $r_{50}$ and $r_{90}$ equal to zero (hence $r_{50}/r_{90}$ is undefined), leaving a set of 29,429.

It was found that, as suspected, the large networks probably were overfitting the Shimasaku et al. training set, since the outputs on running them on the SDSS EDR set were clearly incorrect (e.g. all the same type). Thus the simpler networks were run, using just the concentration parameter, and gave the results in table 4.10. Again further runs should be performed to get a better set of averages, but the overall observation that the mean percentages of elliptical, spiral and irregular galaxies are in the ratio $14\pm 5:86\pm 12:0\pm 0.1$ compares reasonably with the observed values of approximately 20:80:1 at low redshift. (The galaxies in this SDSS EDR set have a mean redshift $z=0.1034 \pm 0.00011$; the $\pm 5, 12$ and $0.1$ come from the sum of errors $\sigma ^2 = \sigma_A^2 + \sigma_B^2 + \ldots$ where $\sigma$ is the absolute error.) The more detailed types in the table are distorted at each end (types 0 and 5, 6 are rarely output), because the use of just the concentration parameter causes the distortion seen in figure 4.4.

\begin{table}
\begin{center}
\caption[Results from applying the networks to the SDSS Early Data Release]{Results from applying the networks to the SDSS Early Data Release. Each network type is run using just the concentration parameter from the 454 Shimasaku et al. set to train. Each is then run (training and classification) on the first 100 galaxies of the SDSS EDR set 10 times, and an average taken, and on the full set once. For a trained network to classify the 29,429 galaxies of the full set takes around 10 minutes, as shown. The networks are all sigmoid with linear output and trained using resilient backpropagation. The single linear neuron is trained for 675 epochs (where its MSE 5th decimal place stayed the same for the first time, as above); the rest were trained for 1000 epochs.}
\gap
{\footnotesize
\begin{tabular}{|c|c|c|c|}
\hline
Network          &Run   &MSE     &Number of each                               \\
                 &      &        &Shimasaku Type                               \\
                 &      &        &                                             \\
                 &      &        &\verb+  E0    S0    Sa    Sb   Sc  Sdm   Im+ \\
                 &      &        &\verb+   0     1     2     3    4    5    6+ \\
\hline \hline                                                                  
1                &100   &1.23384 &\verb+   0    12    47    27   12    2    0+ \\     
8:1              &      &1.10838 &\verb+   0    12    33    40   15    0    0+ \\
8:8:1            &      &1.04274 &\verb+   0    13    29    48   11    0    0+ \\
64:1             &      &1.03261 &\verb+   0     5    36    43   16    0    0+ \\
\hline
1                &29429 &1.23384 &\verb+  48  5272 11658  8603 3511  236  101+ \\    
8:1              &      &1.11082 &\verb+ 311  4938  8578  9297 6220   12   73+ \\
8:8:1            &      &1.04206 &\verb+ 414  2415  9411 14887 2302    0    0+ \\
64:1             &      &1.03261 &\verb+ 582  2375  9566 12059 4780    9   58+ \\
\hline \hline
Mean for the     &      &        &\verb+ 339  3750  9803 11212 4203   64   58+ \\
four 29,429 runs &      &        &                                             \\
\hline
Percentage       &      &        &\verb+ 1.2    13    33    38   14 0.22 0.20+ \\
\hline
Stdev for the    &      &        &\verb+0.76   5.3   4.5   9.8  5.7 0.39 0.14+ \\
four 29,429 runs &      &        &                                             \\
(in percent)     &      &        &                                             \\
\hline
\end{tabular}
}
\end{center}
\end{table}

\subsection{Further Steps}

The next step should be to train the various networks on a subset of the Shimasaku data to assess and hopefully quantify which ones \emph{are} overfitting the data, and then use the best ones which are not doing so on the EDR set. On the other hand, maybe a larger training set is needed, since the Shimasaku set is from the commissioning data, which, like the EDR, is from a narrow band of sky. Thus large scale structures in the region could distort the percentages of galaxy types. (Clusters have a higher percentage of ellipticals and S0's, up to 50\% or more than the mean, and more sparsely populated regions (parts of walls which are not clusters, filaments and voids) have more spirals.)

Other steps would be to vary the learning rate for the actual networks as opposed to just the single neuron (each input weight, layer weight and bias rate must be explicitly set using matrices of the correct connectivity net.inputConnect, net.layerConnect or net.biasConnect (see Appendix B)), and to include the galactic extinction, since the parameters is given in the SDSS EDR set.

Further possibilities are discussed in Chapter 5.

\chapter{Extensions to the Project}

There are many possible ways to extend the project, from specific further work which could be done to more general aims which the community is working towards.

\section{Further Networks and Data}

The very large number of possibilities on M{\sc{atlab}} and on other available programs have not been exhausted by this project. Some possibilities were described in \S 3.7. Many further network programs are available on the web. The FAQ site for the newsgroup comp.ai.neural-nets lists, as of \today , 40 commercial programs and 43 shareware programs, plus numerous source code locations in its parts 5 and 6. The site {\tt http://www.emsl.pnl.g-\\ov:2080/proj/neuron/neural/systems/shareware.html} lists 62 commercial and 59 shareware.

Other examples of programs used for galaxy classification (URLs in references) include LMorpho, and Autoclass (Cheeseman \& Stutz 1996, Goebel et al. 1989). There are also many dedicated codes written by research groups for their particular projects.

Further data would include the digital sky surveys (SDSS, 2dF, etc.), the scanned photographic plates (APM, SuperCOSMOS (Hambly et al. 2001)), and the previously classified galaxy atlases (ESO-LV, RC3, etc.). Of these the new surveys would be the most useful, because of their unprecedented amount, quality and consistency of data.

\section{Further Morphological Schemes}

The networks here have produced Shimasaku types as output. These are based on Hubble's scheme. Clearly other schemes, such as the Yerkes, DDO etc. could be used. However, expert human eyeball classification for a large galaxy sample by a group of independent human experts has been, and only could have been, performed using Hubble's scheme. So other schemes would require checking of a network's reliability by humans, to make sure that it is correctly applying the scheme.

\section{Spectral Schemes}

Classifying galaxies spectroscopically as opposed to morphologically would be the most important general extension to the work in this project. Indeed, in the choices presented as MSc projects it was a possible alternative to this one. Essentially similar work could be carried out. Much literature suggests that spectral classifications are the most meaningful in terms of parameters which can be observed and predicted by numerical or semi-analytic galaxy formation models, and by similar models of galaxy evolution. However, parameters such as the concentration index are equally important and so the best classification systems will probably use both spectral and morphological parameters. This project uses galaxy colours, which are basically integrated stellar spectra, and so could be called spectral. However, the concentration index has been the most significant parameter, hence the 'morphological' in the project title.

\section{Unsupervised Learning}

The two schemes above, and as many meaningful parameters as possible could be fed in (not necessarily all at once) to an unsupervised network to see what patterns it comes up with. The two unsupervised learning methods of nonlinear principal component analysis and the Kohonen Self Organising Map could be used. An example of the latter is Naim et al. (1997), who investigate patterns in parameters for galaxies at moderate redshift.

\section{Other Wavebands and 'Channels'}

The galaxies here were imaged at optical wavelengths. The appearance of galaxies alters considerably at other wavelengths (e.g ultraviolet, infrared), and this could be investigated. An example is the Block et al. (1999) result that that the appearance of galaxies at near infrared wavelengths does not correlate with Hubble type, because galactic dust is much more transparent at these wavelengths.

The electromagnetic spectrum is only one astrophysical 'channel' for which information is available from the sky. Others include neutrinos, gravitational waves or dark matter. Of these, only neutrinos have been detected directly (and then with great difficulty), but each is present in galaxies so should eventually be explained. Indeed, the 'theorist's view' of a galaxy is often that of a large approximately spherical halo of dark matter with a small smudge of luminous matter somewhere in the middle! (The dark matter contains most of the mass, which is important in the theory and simulation of galaxy and large scale structure formation.)

\section{Galaxy Evolution}

The galaxies in this project are essentially at redshift zero (i.e. at $z \sim 0.1$), so the effects of galactic evolution are small. Galaxies at higher redshifts could be investigated, and indeed have been. An example is the detailed investigation of galaxy morphologies within the Hubble Deep Field (Abraham et al. 1996), and many papers have been published suggesting quantitative schemes for extending the Hubble sequence to moderate and high redshifts (e.g. Abraham \& Merrifield 2000). The Hubble sequence has been shown to be inadequate in describing galaxies at redshifts where evolution is significant. Parameters such as the concentration index are useful because they broadly correlate with galaxy type and can be measured to much greater distances than can detailed morphology.

\section{Unusual Galaxies}

The Hubble scheme was designed to include bright, nearby, clearly spiral or elliptical galaxies. However, active galaxies (active galactic nuclei, quasars), interacting galaxies, and the vast numerical majority of small dim galaxies (dwarf ellipticals, dwarf spirals, most irregulars etc., which are therefore not at all 'unusual') are not included. It should be possible to understand these in terms of the parameters the classification schemes show to be meaningful. Other galaxies have recently been detected solely from the 21cm line of neutral hydrogen (HI).

\section{Further Possibilities}

The combination of accurate galaxy types and the breadth, depth and consitency of coverage from the Sloan survey would allow many aspects of galaxies and large scale structure to be investigated in unprecedented detail. For example, the statistics of the structure could be analysed as a function of galaxy type (e.g. clustering strength, luminosity function). These sorts of observations would provide a detailed set of data to constrain simulations and semi-analytic theories. e.g. Benson et al. (2001) quote Sloan for testing $z=0$ and the VIRMOS survey for high redshifts. The next data release for the SDSS is scheduled for January 2003, and the full dataset of 50 million galaxies, 1 million of which will have associated spectra and hence redshifts, should be available by 2005.


\chapter{Conclusions}

The main findings of this project are:

\begin{itemize}

\item Automated classification systems are essential to be able to classify the galaxies in the new digital sky surveys. The largest of these, the Sloan Digital Sky Survey (SDSS), will collect a full dataset of 50,000,000 galaxies in $\pi$ steradians of the northern galactic cap by 2005. Other million-plus datasets exist on photographic plates.

\item The trained networks only perform classification of the galaxies that they have not seen into broad categories, but the intrinsic spread in galaxy properties means that this is sufficient for most purposes. However, with the sheer numbers of galaxies becoming available, what is 'spread' and what is real variation may be distinguishable.

\item The correlation between network type and type found by the human classifiers in Shimasaku et al. (2001) for 454 galaxies from the SDSS commissioning data was around $0.7 \pm 0.1$ for a linear classification method (single linear neuron) using just the concentration index, and improved to around $0.8 \pm 0.1$ for simple networks. These figures also improve by another ~{}0.05 for simple networks compared with the single neuron for the same set of parameters. Larger networks using all the parameters were able to produce correlations arbitrarily close to 1, but these were overfitting the particular training set and would not be able to cope when presented with previously unseen galaxies. A simple network using the most significant parameters (concentration index, one or two colours, and perhaps the profiles) is probably best.

\item Simple networks trained on the Shimasaku et al. (2001) galaxy set were briefly applied to a set of 29,429 galaxies with spectra from the SDSS Early Data Release. They gave mean Shimasaku type percentages corresponding to an elliptical/lenticular:spiral:irregular ratio of $14 \pm 5:86 \pm 12: 0 \pm 0.1$ These are averages from only a small amount of runs, but compare with the observed ratios of \~{} 20:80:1 for galaxies at low redshift. The SDSS data is from a narrow strip of sky, and the networks distort the ends of the distribution, because just the $r_{50}/r_{90}$ concentration parameter was used, as figure 4.4 showed.

\item Morphological galaxy parameters are only a subset of those important for describing galaxies. The most important parameters are those which relate directly to galaxy evolution. These can then be compared with predictions from semi-analytic galaxy formation models. These models connect with the present through models of galaxy evolution, and observations of galaxies at high, medium and low redshift. See e.g. Baugh et al. (2001). Thus the morphological concentration parameter, and the strongly correlated bulge-to-disk ratio are important, as are spectral parameters such as integrated galaxy colour, line strengths and estimated star formation rates. Morphology can often vary significantly when there is not much physical variation (e.g. the rings and S's in galaxy disks which were used in the de Vaucouleurs 3D system). Spectral parameters tend to be more correlated with physical parameters.

\item Supervised networks can never classify galaxies better than the set on which they are trained. This set can be a slight improvement over humans if the set is the mean of several observers, but it will still inevitably be subjective to some extent. Unsupervised classification using principal component analysis (PCA), and a combination of morphological and spectral parameters, is the ideal. Some authors, e.g. Naim et al. (1997) suggest that it may be better simply to think of distributions in parameter space rather than performing classifications. This can then be done at any redshift.

\item In the time available for the project (November 2000--August 2001, full time from June 2001), only a small fraction of the available possibilites could be tried out. Important extensions to the work would be to firstly more fully test the training set available, base the quoted averages on more network runs, and then look in more detail at the SDSS Early Data Release. Other types of network could then be tried, for example unsupervised with PCA, and networks using other other parameters and classification schemes, e.g. spectral. One could then look at galaxies in other wavebands, in particular the infrared where dust in galaxy disks is almost transparent. Finally galaxies at moderate and high redshift, where evolution has become significant, must be described in a quantitative way by any complete classification scheme.

\item The most significant conclusion of the project is probably that the tools are available to create any number of neural networks at a level of sophistication used in the artificial intelligence community, and not just by astronomers. The M{\sc atlab} Neural Network Toolbox is one example of such a toolset, and it is available in a standard form to all. High performance computers would enable more runs and shorter training times than those quoted here. The ideal galaxy classification system would then use the principal component analysis also available in the toolbox, feed the components into an unsupervised network, and classify the galaxies in the SDSS dataset as they become available. The components would probably be directly related to parameters used to describe galaxy evolution, some of which are morphological and some spectral. Trying out all the possibilities in M{\sc atlab} would form a worthwhile PhD or research project.

\end{itemize}

\appendix

\chapter{Acknowledgements}

\section{Official}

\noindent {\bf ADS} \\

\noindent This research has made use of NASA's Astrophysics Data System Abstract Service. \\

\noindent {\bf NED} \\

\noindent This research has made use of the NASA/IPAC Extragalactic Database (NED) which is operated by the Jet Propulsion Laboratory, California Institute of Technology, under contract with the National Aeronautics and Space Administration. \\

\noindent {\bf SDSS} \\

\noindent The Sloan Digital Sky Survey (SDSS) is a joint project of The University of Chicago, Fermilab, the Institute for Advanced Study, the JapanParticipation Group, The Johns Hopkins University, the Max-Planck-Institute for Astronomy (MPIA), the Max-Planck-Institute for Astrophysics (MPA), New Mexico State University, Princeton University, the United States Naval Observatory, and the University of Washington. Apache Point Observatory, site of the SDSS telescopes, is operated by the Astrophysical Research Consortium (ARC). \\
\noindent Funding for the project has been provided by the Alfred P. Sloan Foundation, the SDSS member institutions, the National Aeronautics and Space Administration, the National Science Foundation, the U.S. Department of Energy, the Japanese Monbukagakusho, and the Max Planck Society. The SDSS Web site is http://www.sdss.org/. \\

\noindent Plus, of course, M{\sc atlab} and the astro-ph e-print archive.

\section{And also}

\noindent Help was appreciated from: \\

\noindent {\bf{Jon Loveday (Sussex)}} - project supervisor, thanks for always being around to help when I get stuck, and answering many questions of varying vagueness \\

\noindent {\bf{Kazuhiro Shimasaku (University of Tokyo, Japan)}} - for supplying, pre-publication, a copy of eyeball classifications and parameters for the 456 galaxies in Shimasaku et al. (2001) \\

\noindent {\bf{Ofer Lahav (Cambridge University)}} - for helpful comments and recommending the Bishop (1995) textbook \\

\noindent {\bf{Avi Naim}} - for help with and supplying a copy of the full Naim et al. (1995b) galaxy parameters \\

\noindent {\bf{Stephen Odewahn and Andrew Hopkins (Arizona State and Pittsburg Universities, USA)}} - for help/suggestions with the LMorpho program \\
 
\noindent {\bf{My parents (Sheffield)}} - for paying for this MSc course \\

\noindent {\bf{Sussex University}} - for paying for trips to the 10th November 2000 RAS meeting on 'Surveying the Local Universe' and the 15th January 2001 Cambridge Institute of Astronomy OXCAM3 meeting on 'Revisiting the Hubble Sequence' \\

\noindent {\bf{The other MSc students}} (Diana Hanbury, Aris Kosionidis and Sarah Sharp, in alphabetical order) for being around so that I wasn't the only one! \\

\noindent And everyone around in the computer room who helped with problems. Without exception they were very willing to spend time to help sort things out, and it's been invaluable - thanks! \\

\noindent The project was written using \LaTeXe, using a modified version of the locally available 'thesis.tex' format of Peter Thomas - so thanks to all who helped with my novice's questions on this too.

\chapter{M{\sc atlab} ANN Listing}

This is a sample M{\sc atlab} Neural Network Toolbox listing, written from scratch by the author. Networks used in this project can be obtained by altering lines in the listing, as explained in the comments. The \% signs indicate the comment lines, which are ignored by M{\sc atlab} when the file is run.

{\scriptsize

{\tt

\begin{verbatim}

%************************************************                              *
%------------------------------------------------                              -
%%Classifygeneral:
%%
%%Trained on R input parameters from 'params.dat'
%%file used by Shimasaku et al. 2001:
%%
%%Parameters are combinations of:
%%r50/r90, P_DeV, P_Exp, u*-g*, g*-r*, r*-i*, i*-z*, r*
%%
%%(see tables 5.6 and 5.7)
%%
%%1 output of galaxy Shimasaku Type 0 to 6 (see Shimasaku et al.)
%%
%%Shimasaku Types are eyeball classifications, or 'true'/Shimasaku types
%%Network outputs a corresponding estimate or network type for each galaxy
%%All 454 galaxies are used for training
%%(otherwise only a subset of the information is used)
%%
%%Can also read and classify the file 'spectra2.dat', corresponding
%%to the 29,429 SDSS Early Data release galaxies, once trained on
%%the ones from Shimasaku et al.
%%
%%Uncomment desired network
%%
%% %  = line is commented out
%% %% = line is a comment
%%
%%In this file the network is currently set up to run on parameters 1, 2 and 3,
%%(r50/r90, P_DeV, P_Exp, using a single linear neuron, set to zero initial
%%weights and bias
%%
%%Last update: 27/08/01 Mon
%------------------------------------------------                              -
%************************************************                              *

%------------------------------------------------                              -
%%Load 'params.dat' into a 456 by 24 matrix 'params'
%%The file must be present in the current directory when Matlab is run
%%Each row is one galaxy
%------------------------------------------------                              -

  load params.dat


%------------------------------------------------                              -
%%Set network architecture:
%%
%%Choose an architecture and type the correct
%%network for the run number, as described
%%
%%Note that single 'hardlim' (perceptron),
%%'logsig' and 'tansig' neurons are not used, because:
%%
%%Hardlim is insufficient because the output is 0 or 1 and a continuous output
%%of 0 to 6, or the normalised equivalent, is required
%%
%%Logsig and tansig are not correct for the output layer, which they form
%%in the case of a single neuron, because they would distort the output
%%and compress it to between 0 and 1 and -1 and 1 respectively
%%
%------------------------------------------------                              -

%------------------------------------------------                              -
%%Single linear neurons:
%------------------------------------------------                              -

%%Type 'net = newlin([0 1;0 1; ... ],1) with R [0 1]'s,
%%one for each of the R input galaxy parameters
%%The 0 1 is nominally the input range but it did not
%%make any difference to the results using any range

  net = newlin([0 1;0 1;0 1],1)                                                        

                                
%------------------------------------------------                              -
%%Feedforward backpropagation networks:
%%
%%Purelin is always used as the output layer transfer
%%function so that the output is not distorted
%%S:S:1 can be generalised to S:T:1, etc.
%%but this does not significantly improve performance,
%%and requires further sets of layer weights
%------------------------------------------------                              -

%%S:1 multilayer networks:
%%Type 'net = newff([0 1; 0 1; ... ][S,1]{'fcn','purelin'})

%%where there are R [0 1]'s, S neurons in the first layer
%%(e.g. 8 but important to try others), and 'fcn' is the transfer function
%%(hardlim, purelin, logsig or tansig)

% S=8;

% net = newff([0 1;0 1;0 1],[S,1],{'tansig','purelin'})                               
            

%%S:S:1 multilayer networks:
%%Type 'net = newff([0 1; 0 1; ... ][S,T,1]{'fcn','fcn','purelin'})

% S=8;

% net = newff([0 1;0 1;0 1],[S,S,1],{'tansig','tansig','purelin'})                    


%------------------------------------------------                              -
%%Initialise network here so that any settings
%%set below override a clean network
%%If the weights are explicitly set as opposed to randomised
%%at the start, (which is what Matlab does otherwise) 
%%this then makes the net give the same results for the same 
%%input parameters and settings each time, as it should
%%(Initialisation just before the training section below does not do this)
%------------------------------------------------                              -

  net=init(net);



%************************************************                              *
%------------------------------------------------                              -
%%Set network preferences, including weights and biases
%%Values not explicitly set in the file are defaults (D)
%%
%%The set of specifications given here is long but does
%%completely specify any network used in this project
%%
%%They can be viewed in Matlab by typing their names,
%%when a network has been created       
%%
%%Here " = same as line above
%%
%%
%%
%%Architecture:
%%
%%Settings allow more general connections between the neurons,
%%e.g. last layer to first, etc.
%%These are not employed here                                                   
%%
%%net.numInputs                      - set by 'newlin', 'newff', etc. 
%%                                     i.e. network creation
%%net.numLayers                      - "
%%net.biasConnect                  D - bias set so one connects to each layer:
%%                                     [1] for 1 layer, [1;1] for 2, 
%%                                     [1;1;1] for 3, etc.
%%net.inputConnect                 D - inputs connected to input layer only:
%%                                     [1], [1;0], [1;0;0], etc.
%%net.outputConnect                D - outputs conected to output layer only:
%%                                     [1], [0 1], [0 0 1], etc.
%%net.targetConnect                D - targets connected to output layer only:
%%                                     [1], [0 1], [0 0 1], etc.
%%net.numOutputs                   D - always 1
%%                                     (i.e. 1 galaxy type output by network)
%%net.numTargets                   D - always 1 (one Shimasaku Type per galaxy)
%%net.numInputDelays               D - always 0 (no time dependence in network) 
%%net.numLayerDelays               D - "
%%
%%
%%
%%Subobject Structures:
%%
%%net.inputs{i}                      - i=1, 2, etc., one set for each layer
%%net.inputs{i}.range                - matrix of minimum and maximum values
%%                                     for inputs (set in network creation)
%%net.inputs{i}.size                 - number of inputs
%%                                     (set in network creation)
%%net.inputs{i}.userdata           D - (notes)
%%net.layers{i}
%%net.layers{i}.dimensions           - S, S, ... 1, i.e. number of neurons
%%                                     in each layer (set in network creation)
%%net.layers{i}.distanceFcn        D - dist: function to apply weights to an
%%                                     input to give weighted inputs
%%net.layers{i}.distances            - S by S matrix for each layer
%%                                     (set in network creation) 
%%net.layers{i}.initFcn            D - initialisation function for the layers
%%net.layers{i}.netInputFcn        D - netsum: function to calculate the overall
%%                                     input to a layer by combining the 
%%                                     weighted and biased inputs                                   
%%net.layers{i}.positions            - [0 1 2 3 4 ...] to S-1 for each layer
%%                                     (set in network creation)
%%net.layers{i}.size                 - S S ... 1, i.e. number of neurons
%%                                     in each layer (set in network creation)
%%net.layers{i}.topologyFcn        D - 'hextop' (not used)
%%net.layers{i}.transferFcn          - hardlim, purelin, logsig or tansig
%%                                     for each layer
%%net.layers{i}.userdata           D - (notes)
%%net.outputs{i}
%%net.outputs{i}.size                - set by net.outputConnect
%%net.outputs{i}.userdata          D - (notes)
%%net.targets{i}
%%net.targets{i}.size                - set by net.targetConnect
%%net.targets{i}.userdata          D - (notes)
%%net.biases{i}                                                                 
%%net.biases{i}.initFcn            D - (none) specific initialisation for biases
%%net.biases{i}.learn              D - 1
%%net.biases{i}.learnFcn           D - (depends on training algorithm)
%%net.biases{i}.learnParam           - lr, other learning functions
%%                                     have extra parameters
%%net.biases{i}.size                 - [5], [1]
%%net.biases{i}.userdata           D - (notes)
%%net.inputweights{i,j}              - one set for each connection
%%net.inputweights{i,j}.delays     D - (none) network has no time dependence,   
%%                                     i.e. it is static
%%net.inputweights{i,j}.initFcn    D - (none) specific initialisation
%%                                     for input weights
%%net.inputweights{i,j}.learn      D - 1
%%net.inputweights{i,j}.learnFcn   D - (depends on training algorithm)
%%net.inputweights{i,j}.learnParam   - lr, etc.
%%net.inputweights{i,j}.size         - [S R] size of input weight matrix
%%net.inputweights{i,j}.userdata   D - (notes)
%%net.inputweights{i,j}.weightFcn  D - dotprod (scalar product)
%%net.layerweights{i,j}
%%net.layerweights{i,j}.delays     D - (none) network is static                 
%%net.layerweights{i,j}.initFcn    D - (none) specific initialisation
%%                                     for layer weights
%%net.layerweights{i,j}.learn      D - 1
%%net.layerweights{i,j}.learnFcn   D - (depends on training algorithm)
%%net.layerweights{i,j}.learnParam   - lr, etc.
%%net.layerweights{i,j}.size         - [T S] size of layer weight matrix
%%                                     between layers with S and T neurons
%%net.layerweights{i,j}.userdata   D - (notes)
%%net.layerweights{i,j}.weightFcn  D - dotprod
%%
%%Functions:
%%
%%net.adaptFcn                     D - adaptwb (used for incremental as
%%                                     opposed to batch training)
%%net.initFcn                      D - (depends on training algorithm)          
%%net.performFcn                     - mse (mean square error between network 
%%                                     output galaxy type and Shimasaku Type)
%%net.trainFcn                       - e.g. trainrp
%%                                     (see 'Set Training Parameters' below)
%%
%%Parameters:
%%
%%net.adaptParam                   D - .passes (1 pass through network is the
%%                                     default before the learning algorithm
%%                                     is applied)
%%net.initParam                    D - (none)
%%net.performParam                 D - (none)
%%net.trainParam.epochs              - set below: number of epochs to train for 
%%net.trainParam.goal              D - default is mse=0, usually replaced
%%                                     by .epochs or .min_grad 
%%net.trainParam.lr                  - learning rate: set for single neurons,
%%                                     left on default 0.01 for networks
%%net.trainParam.max_fail          D - used in 'early stopping' training
%%net.trainParam.min_grad          D - gradient value at which to stop training
%%                                     at which to stop training (no value set)
%%net.trainParam.show              D - show mse etc. every .show number
%%                                     of epochs (default is 25)
%%net.trainParam.time              D - stop training after certain
%%                                     processor time (default is infinite)
%%
%%Weight and bias values:
%%
%%net.IW{i,j}                        - can be set below, with indices
%%                                     corresponding to net.inputConnect
%%net.LW{i,j}                        - can be set below, with indices
%%                                     corresponding to net.layerConnect
%%net.b{i}                           - can be set below, with indices
%%                                     corresponding to net.biasConnect                       
%%
%%Other:
%%
%%net.userdata                     D - (notes)
%%
%------------------------------------------------                              -
%************************************************                              *

%------------------------------------------------                              -
%%Set other preferences to zero to clear previous run
%%And show that they are zero
%%If some (e.g. bias3) are not used then they are shown as zero again below,
%%as opposed to causing an error by being undefined
%------------------------------------------------                              -

  iw=0
  lw=0
  bias=0
  bias2=0
  bias3=0
  lr=0 
  ep=0


%-----------------------------------------------                               -
%%Ask for, or leave to default:
%%
%%lr = learning rate (as high as possible without diverging weights)
%%ep = number of epochs (so that MSE asymptote is approximated)
%%
%------------------------------------------------                              -

  lr=input('Learning rate (e.g. 0.001): '); 
  ep=input('Number of training epochs (e.g. 1000): ');


%------------------------------------------------                              -
%%Set weights and bias
%%
%%Uncomment set for appropriate run (table 4.7) and network architecture
%------------------------------------------------                              -

%%To explicitly set the weights and biases and stop them being randomised
%%(most useful when doing quick comparison
%%of parameter combinations with a single linear neuron)


%%Single neurons (no layer weights lw)
%%Type iw=[x y z ...]; with R values for R inputs
%%Usually x=y=z=...=0 as a starting point
%%No layer weights are set

  iw=[0 0 0];                   

  lw=[];

%%S neurons in first layer
%%Type iw=[x1 y1 z1 ...;x2 y2 z2 ...; ...] with R values
%%between each semicolon and S sets of values                                   
%%Again x1=x2=y1=y2= ... =0 is usual
%%Currently S=8 (value of T irrelevant for input weights)

% iw=[0 0 0;0 0 0;0 0 0;0 0 0;0 0 0;0 0 0;0 0 0;0 0 0];                                                  


%%Layer weights for networks with S neurons in first layer
%%Type lw=[x y z ...] for S neurons in each hidden layer
%%(i.e. layers which are not the output layer)

% lw=[0 0 0 0 0 0 0 0];


%%Biases
%%The arrays with more than one zero require S zeros,
%%each separated by semicolons
%%Uncomment the appropriate set and change if necessary                         

%%Single neuron

  bias=[0];

%%S:1

% bias=[0;0;0;0;0;0;0;0];
% bias2=[0];

%%S:S:1

% bias =[0;0;0;0;0;0;0;0];
% bias2=[0;0;0;0;0;0;0;0];
% bias3=[0];


%------------------------------------------------                              -
%%Convert the iw, lw and bias values to the network notation
%%Ones which are zero are weight combinations which are not used,
%%e.g. from 2nd layer to first, etc.
%%They are shown for completeness (the %% lines)
%%Uncomment appropriate set
%------------------------------------------------                              -

%%Single neuron

  net.IW{1}=iw;
  net.b{1}=bias;


%%Multiple neurons

% net.IW{1,1}=iw;
%%net.IW{2,1}=0;

% net.LW{2,1}=lw;
%%net.LW{1,1}=0;
%%net.LW{1,2}=0;
%%net.LW{2,2}=0;

% net.b{2}=bias2;


%------------------------------------------------                              -
%%Set learning rates for weights and biases (all equal to lr, set above)
%------------------------------------------------                              -

%%Single neuron

  net.inputWeights{1}.learnParam.lr=lr;
  net.biases{1}.learnParam.lr=lr;


%%S:1 configuration

% net.inputWeights{1,1}.learnParam.lr=lr;
% net.layerWeights{2,1}.learnParam.lr=lr;
%%Rest are again zero

% net.biases{1}.learnParam.lr=lr;
% net.biases{2}.learnParam.lr=lr; 

%%S:S:1 etc. require further setting from 'iw=0' onwards above,
%%as shown in the matrices for net.biasConnect, net.inputConnect
%%and net.layerConnect


%------------------------------------------------                              -
%%Set training parameters
%%trainFcn's available are: trainb, trainbfg, trainbr,
%%traincgb, traincgf, traincgp, traingd, traingda, traingdm,
%%traingdx, trainlm, trainoss, trainrp, trainscg
%%
%%Some of these have extra parameters
%%See Matlab Neural Network Toolbox site
%%(Chapter 5 in PDF manual) for details of each
%%
%%Single neurons use 'trainwb', i.e. train by weights and bias
%%'trainlm' and 'trainrp' are good for galaxy classification
%%(trainlm for small nets, trainrp for any size)
%------------------------------------------------                              -

% net.trainFcn=trainrp

  net.trainParam.epochs=ep;
% net.trainParam.min_grad=;
% net.trainParam.goal=gl;


%------------------------------------------------                              -
%%Show values before training starts
%%Zeros are shown if not set above
%------------------------------------------------                              -
                                                            
  iw
  lw
  bias
  bias2
  bias3
  lr 
  ep



%************************************************                              *
%------------------------------------------------                              -
%%Training
%------------------------------------------------                              -
%************************************************                              *

%------------------------------------------------                              -
%%Set inputs p and targets t initially to zero to clear previous network run
%%This is not done by the network initialisation function
%%The run only uses the p's corresponding to the desired input galaxy parameters
%%p1 ... p8 correspond to each parameter from table 5.6
%%See table 4.7 and below to see which parameter combinations are run
%------------------------------------------------                              -

  p1=0; %r50/r90
  p2=0; %P_Dev
  p3=0; %P_Exp
  p4=0; %u*-g*
  p5=0; %g*-r*
  p6=0; %r*-i*
  p7=0; %i*-z*
  p8=0; %r*

  t=0;


%------------------------------------------------                              -
%%Show that each is zero in an array rather than on 9 separate lines
%------------------------------------------------                              -

  Inputs_zero=[p1 p2 p3 p4 p5 p6 p7 p8]
  Target_zero=[t]


%------------------------------------------------                              -
%%Read the input data matrix params, from rows 1 to 454
%%(Unneeded p values can be commented out to save processing time)
%%The for loop is ended further below
%%This is the point where the parameters could be normalised
%------------------------------------------------                              -

  for a=1:454;

  p1(a)=params(a,21); 
  p2(a)=params(a,22); 
  p3(a)=params(a,23);
% p4(a)=params(a,11)-params(a,12);
% p5(a)=params(a,12)-params(a,13);
% p6(a)=params(a,13)-params(a,14);
% p7(a)=params(a,14)-params(a,15);
% p8(a)=params(a,13);

  t(a)=params(a,20);


%------------------------------------------------                              -
%%Set target outputs as the values read
%%
%%The value of P is not shown (hence the ;), because it is a LARGE matrix!
%------------------------------------------------                              -

%%Type P=[pa; pb etc.]; to match read parameters
%%e.g. P=[p1; p2; p3]; for run123

  P=[p1; p2; p3];


%------------------------------------------------                              -
%%Train the network on rows st-fin
%%using chosen training algorithm
%%trainlm, trainrp, traingdm etc.
%%
%%If the end were placed after the [net,tr], this would
%%give incremental as opposed to batch training
%------------------------------------------------                              -

  end

  [net,tr]=train(net,P,t);


%------------------------------------------------                              -
%%Use the trained network to classify all the galaxies in the desired data file:
%%
%%params.dat   = Shimasaku et al. 2001
%%spectra2.dat = Sloan Digital Sky Survey Early Data Release
%%               29,429 galaxies with spectra
%%
%%Read relevant columns of loaded matrix rows st-fin and run net on each row,
%%Store results in z
%%
%%err gives difference between trained network's type and true galaxy type
%%Doesn't matter if training set is used again in classification run
%------------------------------------------------                              -

%------------------------------------------------                              -
%%Trained network asks for the rows in the data which require classifying
%%
%%st = row at which to start
%%fin = row at which to finish
%------------------------------------------------

  st=input('Row number to start classifying at (1-454, 830 or 29433): ');
  fin=input('Row number to finish at: ');

    
%------------------------------------------------                              -
%%Read the loaded data file.
%%params.dat is already loaded so it doesn't
%%need to be loaded again if being read here
%%
%%If a set is being classified which is smaller than a previous
%%run's set, Matlab must be restarted to clear z(b), 
%%which will be the size of the previous run
%------------------------------------------------                              -

  for b=st:fin;

%%To classify Shimasaku et al.:

  q1=params(b,21);
  q2=params(b,22);
  q3=params(b,23);
% q4=params(b,11)-params(b,12);
% q5=params(b,12)-params(b,13);
% q6=params(b,13)-params(b,14);
% q7=params(b,14)-params(b,15);
% q8=params(b,13);

%%To classify the SDSS Early Data Release 29,429 set:

% load spectra2.dat

% q1=spectra2(b,17)/spectra2(b,18);
% q2=spectra2(b,20);
% q3=spectra2(b,21);
% q4=spectra2(b,7)-spectra2(b,8);
% q5=spectra2(b,8)-spectra2(b,9);
% q6=spectra2(b,9)-spectra2(b,10);
% q7=spectra2(b,10)-spectra2(b,11);
% q8=spectra2(b,9);

%%Type Q=[qa; qb etc.] to match read parameters
%%e.g. Q=[q1; q2; q3] for run123

  Q=[q1; q2; q3];

  z(b)=sim(net,Q);
  err(b)=z(b)-params(b,20);
  errsq(b)=err(b)^2;

  end



%************************************************                              *
%------------------------------------------------                              -
%%Output statistics from the network
%------------------------------------------------                              -
%************************************************                              *

%------------------------------------------------                              -
%%Weights and bias after training, used to classify the galaxies
%%rmserr = RMS error between network type and true type
%%(equals square root of Matlab's MSE output)
%%stddev = Standard deviation of network types
%%correlation = Spearman's rank correlation coefficient (-1 to 1) 
%------------------------------------------------                              -

  iw
  lw
  bias
  bias2
  bias3

  rmserr=sqrt(mean(errsq))
 
  for c=st:fin;
  t(c)=params(c,20);
  end

  correlation=corrcoef(t,z)



%*************************************************                             * 
%-------------------------------------------------                             -
%%Plot results (requires 'xhost' [machine name] command
%%on local machine if displaying window from a remote machine) 
%-------------------------------------------------                             -
%*************************************************                             *

%-------------------------------------------------                             -
%%Form the 'y=x' line where network output equals the Shimasaku Type
%%
%%x2 y2 etc. are for offset test - makes no difference to correlation
%-------------------------------------------------                             -
 
  for x=1:7;
% x2(x)=x-1;
% y2(x)=x;
  x2(x)=x-1;
  y2(x)=x-1;
  end


%-------------------------------------------------                             -
%%Plot network output versus Shimasaku type
%%
%%'bins' gives the centres of the histogram bins, i.e. Shimasaku types
%%The axes are automatically scaled
%-------------------------------------------------                             -
 
  bins = [0 1 2 3 4 5 6];
 
  plot(t,z, 'bo', x2,y2, 'g-')

  xlabel('Shimasaku Type')
  ylabel('Network Type')
 
  grid on
 
  legend('Results', 'Types equal')


%-------------------------------------------------                             -
%%Hist is a histogram of binned galaxy types (nearest integer)
%%Can compare numbers of each galaxy type: histogram
%%outputs the number of galaxies in each bin
%-------------------------------------------------                             -

% histogram=hist(z(st:fin),bins)
% hist(z(st:fin),bins)         

% xlabel('Shimasaku Type')
% ylabel('Number')



\end{verbatim}

}

}

\addcontentsline{toc}{chapter}{References}
\markboth{{\sl REFERENCES}}{{\sl REFERENCES}}

\chapter*{References}

\noindent {\LARGE References} \\

\noindent Abraham, R. G., Valdes F., Yee H. K. C., van den Bergh, S., 1994, AJ, 432, 75 \\
Abraham, R. G., Tanvir, N. R, Santiago, B. X., Ellis, R. S., Glazebrook, K., van den Bergh, S., 1996, MNRAS, 279, L47 \\
Abraham, R., \& Merrifield, M., 2000, AJ, 120, 2835 \\
Adams, A., \& Woolley, A., 1994, Vistas in Astronomy, Special Issue on Artificial Neural Networks in Astronomy, 38, 273 \\
Baugh, C. M., Benson, A. J., Cole, S., Frenk, C. S., Lacey, C. G., 2001, astro-ph 0103156\\
Bazell, D., \& Aha, D. W., 2001, ApJ, 548, 219 \\
Bishop, C.M., 1995, Neural Networks for Pattern Recognition, Clarendon Press, Oxford \\
Block., D.L., \& Puerari, I., 1999, A \& A, 342, 627 \\
Cheeseman, P. \& Stutz, J., 1996. In: Advances in Knowledge Discovery and Data Mining, eds. Fayyad, U. M., Piatetsky-Shapiro, G., Smyth, P., Uthurusamy, R., AAAI Press/MIT Press
Colless, M. et al., 2001, astro-ph 0106498 \\
de Vaucouleurs, G., 1948, Annales d'Astrophysique, 11, 247 \\
de Vaucouleurs, G., 1959, Handbuch der Physik, 53, 275 \\
de Vaucouleurs, G., de Vaucouleurs, A., 1964, Reference Catalogue of Bright Galaxies, University of Texas, Austin \\
de Vaucouleurs, G., de Vaucouleurs, A., Corwin, H., Buta, R., Paturel, G., Fouqu\'e , P., 1991, Third Reference Catalogue of Bright Galaxies, Springer-Verlag, New York \\
Davis, M., Huchra, J., Latham, D. W., Tonry, J., 1982, ApJ, 253, 423 \\
Doi, M., Fukugita, M., Okamura, S., 1993, MNRAS, 264, 832 \\
Fairall, A.P., 1998, Large Scale Structures in the Universe, Wiley Praxis, Chichester \\
Folkes, S. R., Lahav, O., Maddox, S. J., 1996, MNRAS, 283, 651 \\
Freeman, K. C., 1970, ApJ, 160, 811 \\
Frei, Z., Guhathakurta, P., Gunn, J. E., 1996, AJ, 111, 174 \\
Fukugita, M., et al., 2001, in preparation \\
Gish, H., 1990. In: IEEE Conference on Acoustics, Speech and Signal Processing, p. 1361 \\
Goebel, J., Stutz, J., Volk, K., Walker, H., Gerbault, F., Self, M., Taylor, W., Cheeseman, P., 1989, A\& A, 222, L5 \\
Hambly, N., et al., 2001, astro-ph 0108286, 0108290, 0108291
Hubble, E., 1926, ApJ, 64, 321 \\ 
Hubble, E., 1936, The Realm of the Nebulae, Yale University Press, New Haven, USA \\
Humason, M. L., 1936, ApJ, 83, 18 \\
Lahav, O., et al., 1995, Science, 267, 859 \\Lahav, O., Naim, A., Sodr\'e Jr., L., Storrie-Lombardi, M. C., 1996, MNRAS, 283, 207 \\
Lauberts, A., \& Valentijn, E. A., 1989, The Surface Photometry Catalogue of the ESO-Uppsala Galaxies, ESO \\
Lupton, R., Gunn, J. E., Ivezic, Z., Knapp, G. R., 2001, astro-ph 0101420 \\
Maddox, S. J., Efstathiou, G., Sutherland, W. J., Loveday, J., 1990, MNRAS, 242, 43P \\
Miller, A. S., 1993, Vistas in Astronomy, 36, 141 \\
Morgan, W. W., \& Mayall, N. U., 1957, PASP, 69, 291 \\
Morgan, W. W., 1958, PASP, 70, 364 \\
Naim, A., et al., 1995a, MNRAS, 274, 1107 \\
Naim, A., Lahav, O., Sodr\'e Jr, L., Storrie-Lombardi, M. C., 1995b, MNRAS, 275, 567 \\
Naim, A., Ratnatunga, K. U., Griffiths, R. E., 1997, ApJS, 111, 357 \\
Odewahn, S. C., Windhorst, R. A., Driver, S. P., Keel, W. C., 1996, ApJ, 472, L13 \\
Riedmiller, M., \& Braun, H., 1993, Proceedings of the IEEE International Conference on Neural Networks \\
Sandage, A., 1975. In: Galaxies and the Universe, eds. Sandage, A., Sandage, M., Kristian, J. \\
Schlegel, D. J., Finkbeiner, D. P., Davis, M., 1998, ApJ, 500, 525 \\
Seitter, W. C., 1987. In: Large Scale Structures in the Universe, Proceedings, Bad Honnef, Fed. Rep. of Germany, Springer-Verlag, Berlin\\
Serra-Ricart, M., Calbet, X., Garrido, L., Gaitan, V., 1993, AJ 106, 1685 \\
Shectman, S.A., Landy, S. D., Oemler, A., Tucker, D. L., Lin, H., Kirshner, R. P., Schechter, P. L., 1996, The Las Campanas Redshift Survey, AJ, 470, 172 \\
Shimasaku., K., et al., 2001, astro-ph 0105401 \\
Spiekermann, G., 1992, AJ, 103, 2102 \\
Storrie-Lombardi, M.C., Lahav, O., Sodr\'e Jr, L., Storrie-Lombardi, L. J., 1992, MNRAS, 259, 8p \\
Storrie-Lombardi, M.C., \& Lahav, O., 1994. In: Handbook of Brain Theory and Neural Networks, ed. Arbib, M. A., MIT Press, Boston \\
Stoughton, C., et al., 2001, unpublished (SDSS publication no. 85) \\
Strateva, I., et al., 2001, astro-ph 0107201 \\
Thonnat, M., 1989. In: The World of Galaxies, p. 53, eds. Corwin, H. G. Jr, Bottinelli, L., Springer-Verlag, New York \\
van den Bergh, S., 1960, ApJ, 131, 215 \\
van den Bergh, S., 1998, Galaxy Morphology and Classification, Cambridge University Press \\
Vorontsov-Velyaminov, B.A., Krasnogorskaja, A., Arkipova, V.P., 1962, Morphological Catalog of Galaxies, Vol. 1. (Moscow) \\
Wolf, M., 1908, Pub. Ap. Inst. Konig. Heidelberg, Vol. 3, No. 5. \\
York, D.G., et al., 2000, The Sloan Digital Sky Survey: Technical Summary, AJ, 1579
 
\vspace{1cm}

\noindent {\LARGE Websites} \\

\noindent The URLs are active as of \today. \\

\noindent Autoclass \\
{\footnotesize {\tt{http://ic-www.arc.nasa.gov/ic/projects/bayes-group/autoclass}}} \\

\noindent Frei \& Gunn Galaxy Catalogue \\
{\footnotesize {\tt{http://www.astro.princeton.edu/\~{}frei/catalog.htm}}} \\

\noindent Gene Smith's Astronomy Tutorial \\
{\footnotesize {\tt{http://casswww.ucsd.edu/public/tutorial/Galaxies.html}}} \\

\noindent List of neural network programs available - 1\\
{\footnotesize {\tt{ftp://ftp.sas.com/pub/neural/FAQ.html}}} - see Parts 5 and 6. \\

\noindent List of neural network programs available - 2\\
{\footnotesize {\tt{http://www.emsl.pnl.gov:2080/proj/neuron/neural/systems/shareware.html}}} \\

\noindent LMorpho \\
{\footnotesize {\tt{http://www.public.asu.edu/\verb+~+asusco/documents/lmorpho/INDEX.html}}} \\

\noindent LMorpho Source Code \\
{\footnotesize {\tt{http://www.public.asu.edu/\verb+~+asusco/documents/lmorpho/dist/index.html}}} \\

\noindent Los Alamos e-print Archive (Astro-ph), for preprint papers \\
{\footnotesize {\tt{http://xxx.soton.ac.uk}}} \\

\noindent Matlab Neural Network Toolbox documentation \\
{\footnotesize {\tt{http://www.mathworks.com/access/helpdesk/help/toolbox/nnet/nnet.shtml}}} \\

\noindent NASA Astrophysics Data System (ADS), for published papers in journals \\
{\footnotesize {\tt{http://ukads.nottingham.ac.uk}}} \\

\noindent NASA Extragalactic Database Knowledgebase for Extragalactic Astronomy, for reviews and historical papers \\
{\footnotesize {\tt{http://nedwww.ipac.caltech.edu/level5}}} \\

\end{document}